\documentclass[12pt,nofootinbib,eqsecnum,longbibliography,prd]{revtex4-1}
\usepackage{amssymb}
\usepackage{amsmath}
\usepackage{graphicx}
\usepackage{color}
\usepackage[usenames,dvipsnames]{xcolor}

\newcommand{\private}[1]{}
\newcommand{\comment}[1]{}

\newcommand{\Tr}{{\rm Tr}}

\newcommand{\be}{\begin{equation}}
\newcommand{\ee}{\end{equation}}
\newcommand{\bea}{\begin{eqnarray}}
\newcommand{\eea}{\end{eqnarray}}

\newcommand{\ket}[1]{|#1\rangle}
\newcommand{\bra}[1]{\langle #1|}
\newcommand{\braket}[2]{\langle #1|#2\rangle}

\newcommand{\Poincare}{Poincar\'e\ }

\newcommand{\tsub}[1]{\langle #1 \rangle}
\makeatletter
\def\l@subsection#1#2{}
\def\l@subsubsection#1#2{}
\makeatother
\begin{document}

\title{A CSP Field Theory with Helicity Correspondence}

\author{Philip Schuster}
\email{pschuster@perimeterinstitute.ca}
\affiliation{Perimeter Institute for Theoretical Physics,
Ontario, Canada, N2L 2Y5 }

\author{Natalia Toro}
\email{ntoro@perimeterinstitute.ca}
\affiliation{Perimeter Institute for Theoretical Physics,
Ontario, Canada, N2L 2Y5 }
\date{\today}

\begin{abstract}
We propose the first covariant local action describing the propagation of a {\it single} free continuous-spin degree of freedom.
The theory is simply formulated as a gauge theory in a ``vector superspace'', 
but can also be formulated in terms of a tower of symmetric tensor gauge fields. 
When the spin invariant $\rho$ vanishes, the helicity correspondence is manifest --- 
familiar gauge theory actions are recovered and couplings to conserved currents can easily be introduced. 
For non-zero $\rho$, a tower of tensor currents must be present, of which only the lowest rank is exactly conserved. 
A paucity of local gauge-invariant operators for non-zero $\rho$ suggests that the equations of motion in any interacting theory should be 
{\it covariant}, not invariant, under a generalization of the free theory's gauge symmetry. 
\end{abstract}

\maketitle

\newpage
\tableofcontents
\section{Introduction}\label{Sec:intro}

This paper presents a field theory for a single bosonic continuous spin particle (CSP).  
The theory is defined by a simple free-particle action, which is local and bilinear in a gauge field and can be coupled to an appropriately conserved background current. 
The action is most simply specified using a ``vector superspace'' in which the orientation of an auxiliary 4-vector geometrizes spin.  
This formalism subsumes and simplifies the well-known Schwinger-Fronsdal description of high-spin bosons; CSPs emerge as a simple and natural generalization.
Our focus here is primarily on the bosonic CSP in 3+1 dimensions, but our results apply more generally. 
Of particular interest is the 2+1-dimensional analogue of CSPs, which has a single polarization state with novel properties closely related to the fractional spin of massive anyons.  
Our result completes the program of obtaining local spacetime actions for free particles of every integer spin,
and opens a new door to investigating all long-distance physics compatible with relativity and quantum mechanics. 

\subsection*{Asymptotic States and Quantum Mechanics}

Massive particles in 3+1 dimensions transform in unitary representations of $SO(3)$ and are classified, as is well-known, by their spin-eigenvalue ${\vec J}^2 = S(S+1)$ or, more covariantly, $W^2 = -m^2 S(S+1)$ where $W^{\mu} = \frac{1}{2} \epsilon^{\mu\nu\rho\sigma} J_{\nu\rho} P_\sigma$ and $m^2 = +P^2$, in the mostly-negative metric convention.  An action formalism for massive particles of arbitrary integer spin was formulated by Singh and Hagen in \cite{Singh:1974qz}. 

Massless particles, in contrast, transform in unitary representations of $ISO(2)$, the isometry group of a plane \cite{Wigner:1939cj}.  In the generic faithful representation, $W^2= -\rho^2$ takes an arbitrary negative real eigenvalue (with dimensions of momentum-squared) and the representation comprises infinitely many polarizations.  These are the ``continuous spin'' representations.  The states can always be decomposed into a basis of eigenstates of a ``helicity'' operator ${\vec {\bf J}}\cdot {\hat k}$ for 3-momentum ${\vec k}$.  The two $ISO(2)$ ``translation'' generators (i.e. combinations of transverse boosts and rotations) mix helicities, and can be grouped into raising and lowering operators with $T_\pm |h\rangle = \rho |h\pm 1\rangle$ (see e.g. \cite{CSP1} for a review).  For a given spin-scale $\rho$, 
there are two continuous-spin representations in 3+1 dimensions: a ``bosonic'' one comprising states of all integer helicities, and a ``fermionic'' one comprising states of all half-integer helicities.  In the degenerate case $\rho=0$, each helicity state separates into an independent representation satisfying $(W^\mu - h P^\mu)|h\rangle =0$.  Action formalisms describing free particles of any helicity are known, with the most general high-spin case due to Fang and Fronsdal \cite{Fronsdal:1978rb,Fang:1978wz}.  

Quantum-mechanical consistency places severe constraints on interacting helicity degrees of freedom, suggesting that long-range 
forces in flat space can only be mediated by particles of helicity 2 or less \cite{Weinberg:1964ew,Weinberg:1965rz,Porrati:2012rd}. But this finding crucially relies on the assumption that helicity is exactly invariant under boosts, i.e. $\rho = 0$.
When that assumption is relaxed, quantum consistency permits covariant soft emission amplitudes 
for a CSP that yield finite differential cross-sections and (for small enough $\rho$) viable approximate thermodynamics despite the infinite number of polarizations \cite{CSP1, CSPthermo}. 
Moreover, at energies large compared to $\rho$ or in the non-relativistic limit, these amplitudes approach those of a helicity 0, 1, or 2 particle \cite{CSP2}.  
Thus, relativity and quantum mechanics alone imply that non-trivial CSP physics should  
mirror that of familiar gauge theories and Einstein GR in the $\rho\rightarrow 0$ limit, 
raising the possibility that CSPs might offer consistent models of electromagnetism and gravity that differ from existing descriptions only deep in the infrared. 
Up until now, it has not been possible to investigate these findings from the point of view of a space time field theory for a {\it single} CSP\footnote{Several early papers aimed to describe CSPs with fully covariant fields (rather than a gauge potential) and encountered difficulties \cite{Yngvason:1970fy,Iverson:1971hq,Chakrabarti:1971rz,Abbott:1976bb,Hirata:1977ss}.  A covariant equation of motion for a CSP gauge potential was found in \cite{BekaertMourad}, but was not of Lagrangian form.  Although the authors of this work presented a field theory action recently in \cite{CSP3}, it propagated a continuum of CSP species with \emph{every} spin-scale $\rho$ rather than a single degree of freedom, and its $\rho\rightarrow 0$ limit could not readily be coupled to conserved currents.}, a situation that we are correcting with the formalism presented here. 

\subsection*{Summary of the Free Field Theory}

In addition to familiar coordinate space $x^{\mu}$, the free gauge potential $\Psi(\eta,x)\equiv \psi_0(x)+\eta_{\mu}\psi_1^{\mu}(x)+...$ 
resides in a ``vector superspace'' with four-vector coordinate $\eta^{\mu}$. Only the orientation of $\eta$ is physical, 
and in this space our action is
\be
S=\frac{1}{2} \int d^4 x [d^4 \eta] \left( \delta'(\eta^2+1) (\partial_\alpha \Psi)^2 + \tfrac{1}{2} \delta(\eta^2+1) (\Delta \Psi)^2 \right),
\ee
where $\Delta \Psi \equiv (\partial_\eta.\partial_x+\rho)\Psi$.  $\Psi$ is an analytic function in vector superspace, 
and for such functions the suitably defined $\eta$-integrations yield tensor contractions among the components
of $\Psi$. As we will show in this paper, this action describes a single CSP with spin invariant $\rho$. 
The whole tower of tensors  $\psi_n^{\mu_1...\mu_n}(x)$ in $\Psi$ is needed to describe the tower of massless integer-helicity polarization states present in the spectrum of a CSP. For $\rho\neq 0$, boosts mix the different spin states in exactly the manner expected 
for a particle with $W^2=-\rho^2$.  
Moreover, for non-zero $\rho$ the helicity-$h$ eigenmode cannot be described by a single tensor component in any gauge, but must rather involve infinitely many non-zero tensor components --- in this sense, the whole vector superspace is optional for $\rho=0$ but essential for non-zero $\rho$.

The action is invariant under the gauge transformation
$\delta\Psi_{\varepsilon,\chi} = (\eta\cdot\partial_x - \tfrac{1}{2} (\eta^2+1) \Delta) \epsilon(\eta,x) + (\eta^2+1)^2 \chi(\eta,x)$ where $\epsilon$ and $\chi$ are arbitrary smooth functions. 
In an appropriate component form, the $\epsilon$ gauge transformation reproduces familiar gauge transformations, but with the addition of rank-mixing 
terms proportional to $\rho$, while $\chi$ can be used to fix a double-traceless gauge. 
For $\rho\rightarrow 0$, the action after this partial gauge fixing smoothly and precisely recovers the familiar Fronsdal form, thereby making 
the helicity correspondence conjectured in \cite{CSP2} manifest. Both the gauge potential $\Psi$ and the 
gauge parameters $\epsilon$ and $\chi$ are unconstrained, which differs from the Fronsdal formalism where
double trace degrees of freedom must be removed by hand. Our formalism is naturally endowed with an enhanced gauge symmetry that simply decouples 
all such double traces.

For $\rho=0$, familiar interactions are introduced with a coupling ${S=\int d^4 x [d^4 \eta] \delta'(\eta^2+1)\Psi J}$,
where $J(\eta,x)=J_0(x)+\eta_{\mu}J_1^{\mu}(x)+(\eta_\mu \eta_\nu +\frac{1}{2} g_{\mu\nu})J_2^{\mu\nu} + \dots$. Gauge invariance requires 
$\delta(\eta^2+1)\Delta J=0$, which for $\rho=0$ encodes the conservation of $J_1$ and $J_2$, and forces higher non-derivative currents to vanish.
For non-zero $\rho$, the continuity condition can be satisfied for a tower of component currents obeying
\bea
\partial \cdot J_1 + \rho J_0 &=& 0,\nonumber \\
\partial \cdot J_2^{\mu} + \rho J_1^{\mu} &=& 0, \nonumber \\
\langle\partial \cdot J_3^{\mu\nu} + \rho J_2^{\mu\nu}\rangle &=& 0, \nonumber \\
&...& \nonumber
\eea
where the notation $\langle T \rangle$ denotes the traceless part of $T$. 
Thus, $\Psi$ can be sourced by a current with an exactly conserved lowest-rank current --- for example a conserved 
$J_1^{\mu}$ component would yield a QED-like theory. 
For $\rho\neq 0$, gauge invariance implies that all higher rank currents must 
be non-zero, are not conserved, and naively have $\rho$ suppression increasing with rank. 
This is entirely compatible with helicity correspondence and general constraints on high rank currents in flat space \cite{Weinberg:1964ew,Weinberg:1965rz,Porrati:2012rd}.

\subsection*{Potential Applications}
The action found in the present work does not address the question of CSP-matter interactions, or even CSP self-interactions.  But it simultaneously sharpens the  challenge of constructing such couplings and the motivations to do so.  Like familiar helicity-1 and 2 gauge fields, our CSP field can couple only to appropriately conserved currents. The conservation condition is local and at zeroth order in $\rho$ it is satisfied by scalar, conserved vector, and conserved tensor currents --- echoing in the field theory the S-matrix \emph{helicity correspondence} found in \cite{CSP2}.  
But it is not clear what symmetry the matter sector must possess, in order to satisfy the full $\rho$-dependent continuity condition in a local way (or whether this is even possible). 
Likewise, for non-zero $\rho$, the only gauge-invariant (GI) operators with at most two derivatives and linear in the gauge potential vanish on the equation of motion.
But there are GI operators local in time, but spatially non-local, that recover familiar GI operators in the $\rho\rightarrow 0$ limit and appear causal even for non-zero $\rho$.

At the same time, we can clearly identify two things that are \emph{not} gauge-invariant: a mass term for the CSP field, and a tadpole-like coupling for the trace of the spin-2 mode.  In other words, it may be consistent for a CSP to be scalar-like, in the sense that its high-energy matter amplitudes approach those of a generic scalar particle. Such a particle would look to an effective field theorist like a fine-tuned scalar field, while in fact its masslessness is protected by symmetry! Likewise, even if one can introduce graviton-like CSP-matter couplings, it is not consistent for that CSP to be sourced by vacuum energy, at least not without 
spontaneously breaking the underlying gauge symmetry. An effective field theorist describing the CSP as an Einstein graviton would certainly conclude that it is fine-tuned!

We cannot read too much into these putative violations of effective field theory intuition until we have an interacting theory in hand and can study its quantum consistency.
Indeed, the pessimistic reading is that the above arguments just provide evidence against such interactions existing.  Yet the properties encountered in the free theory and coupling to classical backgrounds hint at just the right kinds of structure to address well known fine-tuning puzzles: a large deformation of the \emph{infrared} degrees of freedom, new symmetries required in the matter sector, and possibly a mild degree of infrared non-locality in the interactions.  We believe this is a strong motivation to resolve the question of CSP couplings to matter one way or the other.

A very different motivation to study continuous spin theories is their potential relevance to 2+1-dimensional physics \cite{panion}.  Continuous-spin particles have analogues in 2+1  dimensions or higher (see \cite{Brink:2002zx} for a classification).  Our action extends straightforwardly to higher and lower dimensions, though we have not fully explored the higher-dimensional state content.  The 2+1-dimensional CSP, unlike its higher-dimensional counterparts, has just two states, related by parity, that transform as $W|\psi\rangle = \pm \rho |\psi \rangle$ with $W=\frac{1}{2}\epsilon^{\mu\nu\rho} J_{\mu\nu} P_\rho$.  To our knowledge, such states have not received any attention in the literature.  

We know of no in-principle obstruction to the emergence of CSP quasiparticles in 2+1-dimensional condensed matter systems.  This would not be without precedent, as massive anyons (with $W|\psi\rangle = \rho |\psi \rangle$, with $\rho=m\cdot s$ arbitrary) play an important role in many such systems (see e.g. \cite{2008AnPhy.323..204S,Wilczek:1990ik}).  From at least one perspective, emergent CSPs would be less surprising than anyons --- while anyon wavefunctions transform non-trivially under a \emph{compact} rotation generator, and therefore must be multi-valued, CSPs merely acquire non-trivial transformations under a non-compact boost + rotation generator.  This is reflected in the simplicity of our action compared to the Jackiw-Nair action needed to represent anyons as elementary fields \cite{Jackiw:1990ka}. 

\subsection*{Outline}

To set the stage, we begin in Section \ref{Sec:helicityS} by summarizing the Schwinger-Fronsdal actions for massless particles of arbitrary integer helicity.
Section \ref{Sec:superspace} introduces ``vector superspace'', and motivates the introduction of gauge symmetries 
to allow unitary representations of helicity degrees of freedom in terms of covariant fields. The resulting action is then shown 
to be equivalent to a sum of Schwinger-Fronsdal actions, but greatly simplified. 
In Section \ref{Sec:cspAction}, we present the action describing a single CSP, showing that the action
propagates massless degrees of freedom with the correct polarization content. We also present the action and equation of motion 
in tensor form, thereby demonstrating that the free theory smoothly and transparently recovers the Schwinger-Fronsdal form as $\rho\rightarrow 0$.
In Section \ref{Sec:physical} we consider couplings of the vector-superspace action to backgrounds currents.  We illustrate how familiar matter-couplings for helicities 0, 1, and 2 can be added to the the vector superspace action with $\rho=0$, and examine the generalized continuity condition for non-zero $\rho$.  In the latter case, gauge-invariance is satisfied by a tower of related currents, of which only the lowest in rank is exactly conserved; we show how appropriately conserved and covariant ansatz currents reproduce the soft CSP emission amplitudes of \cite{CSP1}.
In Section \ref{sec:GIcomments}, we comment on several gauge-invariant and non-invariant operators: the Poincar\'e generators are gauge-invariant, but their spacetime \emph{densities} are not, much like those of gravitational radiation in linearized GR.  We then comment on the absence of simple, local gauge-invariant operators (excluding the equation of motion), and the causality properties of some non-local gauge-invariants.  
We conclude in Section \ref{Sec:conclusion} with a summary, discussion of open problems, and speculation on the physical significance of the paucity of gauge-invariants encountered in Section \ref{sec:GIcomments}.

\section{Review of Massless Bosons in the Schwinger-Fronsdal Formalism}\label{Sec:helicityS}

The Schwinger-Fronsdal formalism describes a propagating helicity-$h$ particle through the dynamics of a rank-$h$ gauge field, which we will denote $\phi^{(h)}$.  The formalism has been extensively reviewed (see e.g. \cite{Sorokin:2004ie,Bouatta:2004kk}), but  we summarize the essential results for completeness and to introduce our notation used to suppress explicit Lorentz indices, which differs slightly from the usual conventions in the high-spin literature. 

Given two tensors $X^{(n)}$ and $Y^{(m)}$ of rank $n$ and $m$, respectively, $X\circ Y$ is the symmetric rank $(n+m)$ tensor obtained by summing over inequivalent orderings of the indices, with no explicit symmetry factor. For example, $(X\circ Y)^{\mu\nu} \equiv X^{\mu}Y^{\nu}+X^{\nu}Y^{\mu}$ for two rank-1 tensors, $(X\circ Y)^{\mu\rho\sigma} \equiv X^{\mu}Y^{\nu\rho}+X^{\nu}Y^{\rho\mu}+X^{\rho}Y^{\mu\nu}$ for a rank-1 with rank-2 tensor and so on. A contraction of two tensors is denoted by $X\cdot Y$, so that $X\cdot Y\equiv X^{\mu_1\dots \mu_n}Y_{\mu_1 \dots \mu_n}$ for two rank-$n$ tensors, $(X\cdot Y)_{\mu}\equiv X^{\nu}Y_{\nu\mu}$ for a rank-1 with rank-2 tensor and so on.  
In actions and other contexts where the end result is clearly a scalar, we will often write contractions of a tensor with itself as $X^2$, as was done above.  However, powers of $\partial_\mu$ or the (mostly-negative) flat metric $g_{\mu\nu}$ are always taken to mean outer products.  
Thus, for a rank-$n$ tensor $X$, the rank-$(n-2)$ object $\partial^2 \cdot X = \partial_\mu \partial_\nu X^{\mu\nu\dots}$ should never be confused with the rank-$n$ object $\Box X=(\partial.\partial) X$.  In addition, $X'=\Tr X=g\cdot X$, $X''=\Tr^2 X = g^2 \cdot X$, etc\footnote{We caution that the notation $X^{(n)}$ is used here to denote the rank of $X$, \emph{not} to denote an $n$'th trace as is sometimes done in the literature.  In addition, our use of a mostly-negative metric leads to some overall sign discrepancies with most of the literature.}.

Using these conventions, we can write the Schwinger-Fronsdal action describing a helicity-h particle in $d$ dimensions as 
\begin{align}
S_h &= (-1)^h \int d^d x \;\tfrac{1}{2} (\partial_\alpha \phi)^2 - \tfrac{s}{2}(\partial \cdot \phi)^2 -\tfrac{s(s-1)}{2} \phi' \cdot (\partial\cdot\partial\cdot \phi) - \tfrac{s(s-1)}{4} (\partial_\alpha \phi')^2  \\ 
& \qquad \qquad- \tfrac{s(s-1)(s-2)}{8} (\partial \cdot \phi')^2 -\phi \cdot J^{(h)} \nonumber\\
& = (-1)^h \int d^d x \; \tfrac{1}{2} (\partial_\alpha \phi)^2 - \tfrac{s(s-1)}{8} (\partial_\alpha \phi')^2 -\tfrac{s}{2} \left(\partial\cdot \phi -\tfrac{1}{2} \partial \circ \phi'\right)^2 -\phi \cdot J^{(h)}.\label{FronsdalEqDform}
\end{align}
The rank-h tensor field $\phi$ is restricted to be double-traceless, $\phi''=0$, and the sign factor  $(-1)^h$ ensures a canonical kinetic term for the physical components with our mostly-negative metric.  
The free action ($J=0$) is then invariant under gauge transformations 
\be
\delta \phi = \partial \circ \varepsilon,  \label{FronsdalGaugeVariation}
\ee 
where $\varepsilon$ is a \emph{traceless} rank-$(h-1)$ tensor ($\varepsilon'=0$),
though in general the Lagrangian changes by a total derivative under gauge variations. 

  The trace condition on the gauge parameter $\varepsilon$ is trivial for helicities $h\le2$ and the double-trace condition on the field $\phi$ is trivial for $h\le 3$.  In fact, for ranks 0, 1, and 2 the action above is equivalent (up to total derivatives) to the Klein-Gordon, Maxwell, or linearized  Einstein actions respectively.
The action remains gauge-invariant when coupled to a current $J(x)$ satisfying $J''=0$ and $\partial\cdot J - \frac{1}{2h+d-6}g\circ \partial\cdot J'=0$  --- in other words,  the traceless part of $\partial\cdot J$ must vanish.  
The equation of motion for $\phi$ is most simply written as
\be
{\cal F} = \Box \phi - \partial \circ \partial \cdot \phi + \partial^2 \circ \phi' = 0,
\ee
though in fact the variation of \eqref{FronsdalEqDform} with respect to $\phi$ yields the equivalent but trace-reversed equation ${\cal F} -\frac{1}{2} g\circ {\cal F'} = 0$.  

\section{To Vector Superspace And Back}\label{Sec:superspace}

Symmetric rank-$n$ gauge potentials can locally encode the propagation of helicity-$n$ radiation. 
To describe a bosonic CSP, for which all integer helicities mix under Lorentz transformations, we must treat all integer spins on an equal footing. 
A natural way to do this, which is quite standard in the high-spin literature, is to introduce a 
new four-vector coordinate $\eta^{\mu}$, and consider fields $\Psi(\eta,x)$ 
that are smooth in $\eta$, i.e. they have Taylor expansions
\be
\Psi(\eta,x) = \sum_n \eta^{\mu_1} \dots \eta^{\mu_n} {\psi^{(n)}(x)}_{\mu_1\dots \mu_n} = \sum_n \eta^{n} \cdot \psi^{(n)}(x), \label{eq:smooth}
\ee
whose coefficients $\psi^{(n)}$ are precisely the rank-$n$ symmetric tensors that encode spin-$n$  degrees of freedom.  The compact notation introduced in the last equality follows the conventions of the previous section. 
Lorentz transformations $\Lambda$ are taken to act on $\eta$ and $x$ while translations $a$ will only act on $x$, i.e. under Poincare transformations,
\be
\Psi(\eta,x) \rightarrow \Psi(\Lambda^{-1} \eta, \Lambda^{-1} x + a), 
\ee 
which implies the usual tensor transformation law of the coefficient functions $\Psi^{(n)}$.  Thus, the \emph{orientation} of $\eta$ encodes spin, while the \emph{scale} of $\eta$ can be absorbed by a field redefinition. There is no dynamics in $\eta$-space --- it is just a useful book-keeping device for compactly manipulating many tensors simultaneously.  Typically, this is done by constructing functions of $\Psi(\eta=0,x)$ and its $\eta$-derivatives, which is equivalent to tensor contraction of the components $\psi^{(n)}$.  
While this book-keeping has facilitated finding equations of motion for CSPs \cite{BekaertMourad}, 
no Lagrangian equations of this form have ever been found.

The new approach that we introduce here, which we refer to as ``vector superspace,'' is to embrace the \emph{geometry} of $\eta$-space.  Though the actions we consider are superficially quite similar to those found in \cite{CSP3}, that work considered rather singular functions of 
$\eta$, while here we will consider smooth fields \eqref{eq:smooth}.  The simplicity of the actions thus obtained is remarkable.  
The sum of free Schwinger-Fronsdal actions for integer helicities in four dimensions, usually written as a sum of five terms
for each tensor field $\psi^{(n)}$, can be recast as an integral in $\eta$-space localized to the neighborhood of the unit space-like hyperboloid: 
\be
 \int [d^4\eta] \delta'(\eta^2+1) \tfrac{1}{2}(\partial_\mu \Psi(\eta,x))^2 +\tfrac{1}{4}  \delta(\eta^2+1) (\partial_\eta.\partial_x \Psi)^2.
\label{eq:EtaSpaceSF}
\ee
The notation $[d^4\eta]$ denotes that the integral should be regulated by analytic continuation and appropriately normalized as explained below, and $\delta'(a) = \frac{d}{da} \delta(a)$.   
The generalization to a CSP with nonzero $\rho$ is 
\be
 \int [d^4\eta] \delta'(\eta^2+1) \tfrac{1}{2}(\partial_\mu \Psi(\eta,x))^2 +\tfrac{1}{4}  \delta(\eta^2+1) ((\partial_\eta.\partial_x  + \rho) \Psi)^2,\label{CSPaction1}
\ee
which possesses a $\rho$-deformed gauge invariance.  

The remainder of this section introduces the key features of the integer-helicity action \eqref{eq:EtaSpaceSF} and demonstrates its equivalence to a sum of Schwinger-Fronsdal actions; the bosonic CSP generalization \eqref{CSPaction1} will be considered in detail in the next section.  Sections \ref{Ssec:deltaFunction} and \ref{Ssec:bottomUpGauge} motivate the basic structure of the action --- its localization to an $\eta$-space hyperboloid and the need for a gauge symmetry to encode the $ISO(2)$ little group through the hyperboloid's $SO(3,1)$ symmetry.  Section \ref{Ssec:tensorDecomposition} explicitly constructs the Schwinger-Fronsdal action and equations of motion from a component decomposition of \eqref{eq:EtaSpaceSF}.  Some computational details are deferred to Appendix \ref{App:tensorIntegrals}. 
Though we focus on 3+1-dimensional theories throughout most of this paper, the results in the Appendix suggest simple generalizations to higher dimensions, as well as to 2+1 dimensions as discussed in \cite{panion}.

\subsection{Spin as Superspace Orientation and the Delta Function}\label{Ssec:deltaFunction}
We have already noted the qualitative idea that the orientation of $\eta$ should geometrize the notion of spin, while its magnitude, which could be absorbed into a field redefinition of the $\psi^{(n)}(x)$, is unimportant.  This idea points to a restricted class of Lorentz-invariants built out of $\Psi$.  
For example, we cannot simply integrate products of 
$\eta$-space functions to form invariants, such as $\int d^4\eta \Psi_1(\eta)\Psi_2(\eta)$, as this is sensitive to the profile of $\Psi$ as we re-scale $\eta$, rather than just the orientation.  Moreover, such integrals are not well-defined for smooth wave-functions $\Psi$.

The simplest invariants that are both well-defined and only sensitive to the $\eta$-space orientation-dependence  
of fields are integrals along hypersurfaces.
To guarantee Lorentz-invariance of the resulting action, we focus on the hyperboloids $\eta^2+\alpha^2=0$, 
with Lorentz invariant integrals of the form
\be
\int d^4\eta \delta(\eta^2+\alpha^2)F[\Psi(\eta)], \label{deltaIntegral}
\ee
for some functional $F$. The magnitude of any non-zero $\alpha$ can be absorbed into a field redefinition of $\Psi$, so without loss of generality we take $\alpha=+1$, which corresponds to unit \emph{space like} norm of $\eta$ in our mostly-negative metric convention.  
Integrals of this form still appear to be ill-defined --- for example, if $F[\Psi] =1$ then \eqref{deltaIntegral} is simply the volume of the hyperboloid, which is clearly divergent.  But integrals of the form \eqref{deltaIntegral} are uniquely defined by analytically continuing $\eta$ to a Euclidean-signature variable $\bar\eta$ defined by $(\bar\eta^0,\bar\eta^i) = (i \eta^0, \eta^i)$.
Upon continuation, \eqref{deltaIntegral} becomes an integral of a smooth function over the compact 3-sphere $\bar\eta^2=1$, which is convergent.  
Alternatively, one can define \eqref{deltaIntegral} directly in Minkowski space using only its symmetry properties, up to an overall normalization. 
Other equally well-defined invariants can be formed, for example by integrating over derivatives of $\delta$-functions.  These generalizations 
are discussed more fully in Appendix \ref{App:tensorIntegrals}.

Of course, integrals such as \eqref{deltaIntegral} can equally well be thought of as functionals of the tensor components $\psi^{(n)}(x)$.  
To make contact with this decomposition, it is useful to introduce the generating function
$G(w)\equiv \int d^4\eta \delta(\eta^2+\alpha^2)e^{-i\eta\cdot w} = 2 J_1(x)/x|_{x = \sqrt{-w^2}},$ in terms of which (see also Appendix \ref{App:tensorIntegrals}) 
\be
\int [d^D \eta] \delta(\eta^2+ 1) F(\eta) = \left[ F(i \partial_w) G(w)\right]_{w=0} = \left[ G(i \partial_\eta) F(\eta)\right]_{\eta=0} \equiv F*G.\label{FstarGtext}
\ee
But the geometric form has at least two advantages: the action is quite simple, and many important results can be demonstrated using simple algebraic manipulations or integrations (see e.g. the derivations of the Pauli-Lubanski invariant for a CSP in \S\ref{Ssec:MassAndSpin} and of the orthonormality of basis states in App. \ref{Sapp:orthonormality}).  

\subsection{Gauge-Invariance in Superspace From the Bottom Up}\label{Ssec:bottomUpGauge}

The physical motivation for keeping track of only the orientation of $\eta$ is that it should encode spin -- this is why 
the actions in vector superspace are supported in a neighborhood of a hyperboloid. 
In fact the hyperboloid's symmetry group $SO(3,1)$ is the smallest group that contains an $ISO(2)$ subgroup \emph{and} is related by analytic continuation to a compact group (allowing the definition of surface-integrals by analytic continuation).   For a given null momentum $k$, the little group is spanned by the three Lorentz generators that leave $\eta \propto k$ invariant. 

To specify the generators more explicitly, we introduce a set of null frame vectors $k,q,\epsilon_{\pm}$ with $q\cdot k=1$, $\epsilon_{+}\cdot\epsilon_{-}=-2$, and all other products among frame vectors vanishing\footnote{This normalization convention differs from that of \cite{CSP1,CSP2,CSP3}, where we took $\epsilon_+.\epsilon_- = -1$.  The unconventional normalization used here implies $W^2 = -T_+ T_-$.}.  The three little-group generators are 
\bea
R &=& \frac{1}{2}(\eta\cdot\epsilon_-\epsilon_+\cdot\partial_{\eta}-\eta\cdot\epsilon_+\epsilon_-\cdot\partial_{\eta})  \label{eq:helicityRot} \\
T_{\pm} &=& i(\eta\cdot k\,\epsilon_{\pm}\cdot\partial_{\eta}-\eta\cdot\epsilon_{\pm}\,k\cdot\partial_{\eta}).
\eea
Here, $R$ reduces to the familiar helicity operator if we choose $\epsilon_{\pm}^0=0$, while $T_{\pm}$ are 
the commuting translations (more precisely,$T_+ + T_-$ and $i(T_+ - T_-)$ are the Hermitian translation generators of $ISO(2)$).  The algebra $[R,T_{\pm}]=\pm T_{\pm}$ can be trivially verified from the above%
\footnote{Upon analytically continuing $\eta^0\rightarrow i\eta^0$, $R$ remains Hermitian provided $\epsilon_{\pm}^0=0$, while $T_{\pm}$ continue to \emph{complex} linear combinations of Hermitian $SO(4)$ generators.}.

Although these differential operators realize $ISO(2)$, smooth wave-functions $\Psi(\eta,k)$ do not realize its unitary helicity-$h$ representations, except for the trivial $h=0$ representation. Helicity representations are eigenstates of $R$ annihilated by both $T_{\pm}$. 
But for an eigenstate of $R$, the best we can do is construct a wave-function annihilated by \emph{one} of $T_\pm$.  
For example, $\Psi_h \sim (\eta\cdot\epsilon_+)^h$ is annihilated by $T_+$ but has $T_- \Psi_h \sim h\, \eta\cdot k\, (\eta\cdot\epsilon_+)^{h-1}$ (of course the antisymmetric field strength $F^{\mu\nu}$ and its higher-spin generalizations are annihilated by $T_\pm$, but because $\Psi(\eta,x)$ decomposes into \emph{symmetric} tensors, they can be obtained only as gradients of $\Psi$, consistent with the above).  
Moreover, the functional form of $\Psi_h$ is not fully specified by its helicity eigenvalue; any function $\Psi \sim (\eta\cdot\epsilon_+)^h f(\eta.k, \eta.q)$ is also an $R$-eigenstate.  

Both of the above difficulties are resolved simultaneously by imposing a ``transverse'' requirement on the wave-function and positing a gauge redundancy.  
For example, the transverse condition $k\cdot\partial_{\eta}\Psi=0$ rules out $\eta.q$-dependence of $\Psi$, while a gauge redundancy $\Psi \simeq \Psi + \eta\cdot k\, \varepsilon(\eta,x)$ ensures that $T_- \Psi_h$ is pure gauge, as is the $\eta.k$-dependence of $\Psi$.  
But these two requirements are actually incompatible --- generic gauge variations $\delta \Psi = \eta\cdot k \varepsilon(\eta,x)$ violate the transverse condition!

Fortunately, our physically motivated assumption that only the behavior of $\Psi$ on the surface $\eta^2+1=0$ matters resolves this bind --- for consistency, we should have demanded the weaker transverse condition 
\be
\delta(\eta^2+1) k\cdot\partial_{\eta}\Psi=0 \label{eq:goodTransverse}
\ee
 and the surface-localized gauge equivalence $\delta(\eta^2+1) \Psi \simeq \delta(\eta^2+1)(\Psi + \eta\cdot k \varepsilon(\eta))$.    
Back in position-space, the covariant equations 
$\delta^{\prime}(\eta^2+1)\Box_x\Psi(\eta,x)=0$ and $\delta(\eta^2+1)\partial_x\cdot\partial_{\eta}\Psi(\eta,x)=0$, 
with a gauge redundancy $\delta(\eta^2+1)\Psi\simeq \delta(\eta^2+1)(\Psi + \eta\cdot \partial_x \varepsilon(\eta,x))$ 
realize all the helicity representations --- solutions to these equations decompose (on the $\delta$-function support) 
into helicity eigenstates proportional to $(\eta.\epsilon_\pm)^{|h|}$ plus gauge.  The derivative of a delta function, $\delta'(x)=\frac{d}{dx}\delta(x)$, was needed in the first equation because the transverse condition involves a derivative of $\Psi$, and therefore depends on $\Psi$ in a first neighborhood of $\eta^2+1=0$.

What can we say about the gauge transformation of $\Psi$ itself, away from the 
$\eta^2+1=0$ surface? 
The most general gauge transformation on $\Psi$ that reduces to the above on the $\eta^2+1=1$ surface,  
$\delta_{\varepsilon,\tilde\varepsilon}\Psi=\eta\cdot k \,\varepsilon(\eta) + (\eta^2+1)\tilde{\varepsilon}(\eta)$,
does not preserve the transverse condition $\delta(\eta^2+1) k\cdot\partial_{\eta}\Psi=0$.  
It is easy to verify that the transverse condition \emph{is} invariant under the restricted gauge transformation
\be
\delta\Psi_{\varepsilon,\chi} = (k \cdot \eta - \tfrac{1}{2} (\eta^2+1) k \cdot \partial_\eta) \epsilon(\eta,x) + (\eta^2+1)^2 \chi(\eta,x), \label{eq:fullGaugeRedundancy}
\ee
and that this redundancy suffices to guarantee that $\delta(\eta^2+1) \Psi(\eta,x)$ realizes the integer-helicity representations.
Strictly speaking, the subset of $\epsilon$ transformations with $\epsilon(\eta,x) = (\eta^2+1) \xi(\eta,x)$ for some $\xi$ are entirely equivalent to the gauge-transformations generated by $\chi$.  But it will prove convenient to include both explicitly. 

From these observations, it is easy to discover that the covariant equation of motion 
\be
\delta^{\prime}(\eta^2+1)\Box_x \Psi- \tfrac{1}{2}\partial_x\cdot\partial_{\eta}(\delta(\eta^2+1)\partial_x\cdot\partial_{\eta}\Psi)=0 \label{eq:etaEOM}
\ee
is gauge-invariant under \eqref{eq:fullGaugeRedundancy} with $\Psi$ and $\varepsilon$ unconstrained, and precisely reduces to the on-shell form once 
we specialize to the transverse $\delta(\eta^2+1) \partial_x\cdot\partial_{\eta}\Psi=0$ gauge (we call this ``harmonic gauge''). 

To see geometrically how this equation of motion encodes a tower of helicities, it is useful to fully fix gauge by taking $\Psi(\eta,x)$ to satisfy $\partial_\eta^2 \Psi = \nabla_\eta.\nabla_x \Psi = \partial_\eta^0 \Psi = 0$ (all three conditions can only be satisfied simultaneously on the support of the equation of motion; we defer the proof that such a gauge can be reached to \S\ref{Ssec:planewave}).  In this gauge, the physical plane-wave solutions with momentum ${\vec k}$  are specified by a harmonic function on the spatial 2-plane in $\eta$-space transverse to ${\vec k}$ (e.g. the plane $\eta^0 = \vec\eta.\vec k = 0$); the helicity rotation operator \eqref{eq:helicityRot} simply rotates this plane.  
Identifying the 2-plane with the complex $z$-plane, the helicity operator is $z\partial_z - \bar z \bar\partial_z$ and the harmonic condition $\partial_z \bar \partial_z \Psi(z)=0$ restricts $\Psi$ to be a sum of holomorphic and anti-holomorphic parts, which can be decomposed as a sum of monomials $z^h$ and $\bar z^h$ with helicity $\pm h$ respectively.  The physical data, i.e. coefficients of each monomial in a series expansion for $\Psi(z,\bar z)$, can indeed be obtained by integrating $\Psi(z,\bar z)$ against $z^h$ (or $\bar z^{|h|}$) over the circle defined by the plane's intersection with the hyperboloid $\eta^2+1=0$.  In fact, this identification of states with harmonic functions of $z$ will persist in the generalization of \eqref{eq:etaEOM} to CSPs with $\rho \neq 0$, with $\Psi$ acquiring a non-trivial (but fully determined up to gauge) dependence on a direction transverse to the $z$-plane. 

\subsection{The Fronsdal Formalism Simplified in Vector Superspace}\label{Ssec:tensorDecomposition}
By demanding that spin be realized geometrically and covariantly in vector superspace,
we derived the unified equation of motion \eqref{eq:etaEOM} for massless particles of integer helicity.
This equation of motion is the variation of the quadratic action 
\be
S=\frac{1}{2} \int d^4 x [d^4 \eta] \left( \delta'(\eta^2+1) (\partial_\alpha \Psi)^2 + \tfrac{1}{2} \delta(\eta^2+1) (\Delta \Psi)^2 \right),\label{DdimAction}
\ee
where $\Delta \Psi \equiv \partial_\eta.\partial_x\Psi$ and the notation $[d^D \eta]$ denotes the regulated integration over $\eta$-space as defined in \S\ref{Ssec:deltaFunction} and App.~\ref{App:tensorIntegrals}.
The action \eqref{DdimAction} (but not the Lagrangian) and the equation of motion \eqref{eq:etaEOM} are invariant under the gauge transformations \eqref{eq:fullGaugeRedundancy}.

It is not yet obvious that the simple action  \eqref{DdimAction} is equivalent to the Schwinger-Fronsdal tensor actions. To see this connection, we will need to carry out the $\eta$-integration explicitly.  In terms of the generating functional $G'(w) = J_0(\sqrt{-w^2})$ (see Appendix \ref{App:tensorIntegrals}), we can write the action as
\be
S  =  \int d^4 x \left[ \tfrac{1}{2} G'(i \partial_\eta) (\partial_\alpha \Psi)^2 + \tfrac{1}{4} G(i \partial_\eta) (\Delta \Psi)^2 \right]_{\eta=0}. \label{StarAction}
\ee
Evidently, although the action \eqref{StarAction} is quadratic in fields, it is not rank-diagonal when we expand $\Psi$ into components as in \eqref{eq:smooth} --- for example, the degree-$2n$ term in the Taylor expansion of $G'$ provides a rank-diagonal kinetic term for the field $\psi^{(n)}$, but also mixes the ``scalar'' $\psi^{(0)}$ with the maximal trace of $\psi^{(2n)}$.  
The rank-mixing is related to the fact that tensor components in \eqref{eq:smooth} were defined via Taylor expansion about $\eta=0$, while the action is supported at non-zero $\eta$.  To remove the rank-mixing, we must decompose $\Psi(\eta,x)$ into tensors in a different way.  

In this section we present a simple decomposition that recovers a sum of Fronsdal actions, 
but is only rank-diagonal after partially fixing gauge so that the tensor components are double-traceless.  
An example of a different decomposition, which removes rank-mixing to all orders in traces while maintaining the full gauge symmetry \eqref{eq:fullGaugeRedundancy}, is discussed in Appendix \ref{App:unconstrainedRankDiagonal}.

The nature of the Fronsdal fields' gauge transformations  $\delta\phi^{(n)} = \partial \circ \epsilon^{(n-1)}$ for traceless $\epsilon$ will be our guide to decomposing $\Psi$ into Fronsdal-like components.
We can parametrize arbitrary gauge transformations in terms of tensor components as
\bea
\epsilon(\eta,x) &= &\sum_{n} (n+1) c_{n+1} \eta^n \tilde \epsilon^{(n)}(x)
  = c_1 \tilde \epsilon^{(0)} + 2 c_2 \eta^{\mu}\tilde \epsilon_\mu^{(1)} + 3 c_3 \eta^\mu \eta^\nu \tilde \epsilon_{\mu\nu}^{(2)} +\dots, \label{epsilonDecomposition} \\
\chi(\eta,x) &= &\sum_{n} \eta^n \chi^{(n)}(x) = 
  = \chi^{(0)} + \eta^{\mu}\chi_\mu^{(1)} + \eta^\mu \eta^\nu \chi_{\mu\nu}^{(2)} +\dots, \label{chiDecomposition}
\eea
with the numerical factors $c_n = 2^{n/2}$ are chosen for later convenience.  We can further decompose each $\tilde \epsilon^{(n)} = \epsilon^{(n)} + g\circ \xi^{(n-2)}$ where $\epsilon^{(n)}$ is traceless.  In fact, as was mentioned already in the vector-superspace language, this 
way of parametrizing gauge transformations is redundant. 
For example, any transformation of the form $\tilde \epsilon^{(2)} \propto g \circ \xi^{(0)}$ is equivalent to one generated by 
$\tilde \epsilon^{(0)}\propto \xi^{(0)}$. Likewise, a transformation with $\tilde \epsilon^{(3)} \propto g \circ \xi^{(1)}$ is equivalent to one generated by 
$\tilde \epsilon^{(1)}\propto \xi^{(1)}$ {\it and} $\chi^{(0)}\propto \partial\cdot \xi^{(1)}$.
In general, gauge transformations generated by $\tilde \epsilon^{(n)}\propto g\circ \xi^{(n-2)}(x)$ are equivalent to 
$\tilde \epsilon^{(n-2)}\propto \xi^{(n-2)}$ with $\chi^{(n-3)} \propto \partial\cdot \xi^{(n-2)}$.
Thus, starting at rank 2, the trace of $\tilde \epsilon^{(2)}$ is redundant with a lower rank gauge transformation. 
At rank 3, the trace of $\tilde \epsilon^{(3)}$ is redundant with a lower rank $\tilde \epsilon^{(1)}$ and $\chi^{(1)}$. 
By induction, the trace of $\tilde \epsilon^{(n)}$ will in general be redundant with lower rank traceless 
$\tilde \epsilon$ and $\chi$.
Thus, it suffices to consider the gauge transformations generated by traceless $\epsilon^{(n)}$ and arbitrary $\chi^{(n)}$.  

The gauge variation induced by the traceless gauge parameters $\epsilon^{(n)}$ can be written as 
\bea
\delta \Psi(\eta,x)  & = & \sum_{n\ge1}  c_n \left(\eta^{(n)} - \tfrac{1}{2} (\eta^2+1) g \circ \eta^{(n-2)}\right) \partial \circ \epsilon^{(n-1)}\\
& = & c_1 \eta^{\mu}\partial_\mu \epsilon^{(0)} + c_2 \left(\eta^\mu \eta^\nu - \tfrac{1}{2} g^{\mu\nu} (\eta^2+1)\right) \partial_{(\mu}  \epsilon^{(1)}_{\nu)}  \nonumber \\
& & \qquad + c_3 \left(\eta^\mu \eta^\nu \eta^\rho - \tfrac{1}{2} (g^{\mu\nu}\eta^{\rho} + g^{\mu\rho}\eta^\nu + g^{\rho\nu}\eta^\mu)  (\eta^2+1)\right) \partial_{(\mu} \epsilon^{(2)}_{\nu\rho
)} +\dots \nonumber,
\eea
where $\partial_{(\mu} \epsilon_{\nu \dots )}$ denotes the sum (without a combinatoric factor) over all permutations of the indices $\mu,\nu,\dots$.  The suggestive resemblance to the gauge transformation \eqref{FronsdalGaugeVariation} of the Fronsdal fields motivates the decomposition of $\Psi(\eta,x)$ into tensor components as 
\bea
\Psi(\eta,x) 
& = & \sum_{n}  c_n \left( \eta^n - \tfrac{1}{2} g\circ \eta^{n-2} (\eta^2+1) \right) \phi^{(n)},\label{FronsdalDecomposition}
\eea
which transform as $\delta\phi^{(n)} = \partial \circ \epsilon^{(n-1)}$.
Up to now, the fields $\phi^{(n)}$ are unconstrained --- in particular, they do not satisfy a double-trace condition.  But each of the remaining $\chi^{(n)}$ gauge redundancies gives us precisely enough freedom to fix a gauge where $\phi^{(n+4)}$ is double-traceless, i.e. ${\phi^{(n+4)}}''=0$.

To summarize our results: we can parametrize all inequivalent gauge transformations by a sequence of traceless $\epsilon^{(n)}$ and unconstrained $\chi^{(n)}$.  If we decompose the vector-superspace field $\Psi(\eta,x)$ a la \eqref{FronsdalDecomposition} in terms of tensor components $\phi^{(n)}(x)$, we can always reach a gauge where the $\phi^{(n)}$ are double-traceless.  This requirement fully fixes the $\chi^{(n)}$ gauge symmetries leaving only the traceless $\epsilon^{(n)}$'s, which transform $\psi^{(n)}$ in precisely the same manner as \eqref{FronsdalGaugeVariation}, i.e. $\delta \psi^{(n)} = \partial \circ \epsilon^{(n)}$.  

With this decomposition in hand, it is straightforward to verify that the action \eqref{DdimAction} is equivalent to a direct sum of Schwinger-Fronsdal actions for all $n$.  The coefficients $c_n$ have been chosen so that the component fields $\phi^{(n)}$ are canonically normalized in 4 dimensions.  
For example, to expand the term $\int [d^4 \eta] \tfrac{1}{4} \delta(\eta^2+1) (\Delta \Psi(\eta,x) )^2$ in components, we first note that  
\bea
\Delta \Psi(\eta,x) = \partial_\eta \cdot \partial_x \Psi(\eta,x) &= &\sum_{n\ge 1} n c_n \eta^{n-1} D^{(n-1)}(x)+ (\eta^2+1)\cdots,\\
&& \quad D^{(n-1)} \equiv \left(\partial\cdot \phi^{(n)} - \tfrac{1}{2} \partial \circ {\phi^{(n)}}'\right),
\eea
where the terms proportional to $(\eta^2+1)$ do not contribute to the action because of the $\delta$-function.  
Moreover, $D^{(n-1)}$ is traceless, and is therefore annihilated by $\partial_{\eta}^2$.
 Thus, upon expanding the generating function $G(i \partial_\eta)$ as a power series in $\partial_{\eta}^2$, the only surviving terms in 
 \eqref{StarAction} are those where each $\partial_\eta^2$ acts as $\partial_\eta^\mu \Delta \Psi {\partial_\eta}_\mu \Delta \Psi$, i.e. where two $D$-tensors of the same rank are fully contracted into each other:
\bea
\int [d^4 \eta] \delta(\eta^2+1) (\partial_\eta \cdot \partial_x \Psi(\eta,x) )^2& =& G(i \partial_\eta) \left(\sum_{n\ge 1} 2^{n/2}\,n\, \eta^{n-1} D_{(n-1)}\right)\\
&& \qquad \left(\sum_{m\ge 1} 2^{m/2}\, m\, \eta^{m-1} D_{(m-1)}\right) \bigg|_{\eta=0} \label{firstLine} \\
& = & \sum_n \tfrac{n}{2} D^{(n-1)}(x)\cdot D^{(n-1)}(x),
\eea
which precisely matches the third term in the Schwinger-Fronsdal action \eqref{FronsdalEqDform}.  The final result follows from Taylor expanding the Bessel function in $G$ and a bit of careful counting --- the factor of $n/2$ arises from the product of the Taylor coefficient in the Bessel function, the $2^p p!^2$ combinatoric factor from all the ways that $(\partial_\eta)^{2p}$ can act on the term in \eqref{firstLine} with $n=m=p$, and the explicit factors $(2^{n/2}\,n)\times (2^{m/2}\,m)$.  Completely analogous calculations show that the first two terms of \eqref{FronsdalEqDform} can be obtained by expanding the $\delta'$ term in the vector superspace action \eqref{DdimAction}.

This construction demonstrates that we can view the vector superspace action \eqref{FronsdalEqDform} as a sum over helicities of helicity-$n$ Schwinger-Fronsdal actions, with the tensor fields grouped together in a convenient way.  Conversely, we can view the Schwinger-Fronsdal action for a single helicity as arising from a partial gauge-fixing of the vector superspace action, with the field $\Psi$ restricted to a particular spin sector.  It would be interesting to explore whether this repackaging of the integer-helicity Schwinger-Fronsdal actions in a simple geometric form may generalize to high-spin theories in deSitter and anti-deSitter backgrounds, and to other formulations of high-spin theories \cite{Bouatta:2004kk,Vasiliev:1988xc,Vasiliev:1988sa,Vasiliev:2004qz}, but we do not pursue these questions here.  Instead, we take the $\eta$-space action \eqref{DdimAction} --- which propagates one state of every integer helicity and treats all spins on an equal footing --- as a starting point for building a gauge theory for a free CSP.

\section{Theory of a Single CSP}\label{Sec:cspAction}
We have seen that the gauge potential $\Psi$ with dynamics governed by \eqref{DdimAction} describes
a tower of free helicity-$h$ degrees of freedom, one for each integer helicity value. This is a natural starting point for describing 
a CSP with $\rho\neq0$, as we expect such an object to involve a propagating tower of integer-spaced helicity-like degrees of freedom
that mix under boosts. In fact, a trivial generalization of \eqref{DdimAction} does the job. 

The action for a CSP with nonzero spin invariant $\rho$ is
\be
S=\frac{1}{2} \int d^D x [d^D \eta] \left( \delta'(\eta^2+1) (\partial_\alpha \Psi)^2 + \frac{1}{2} \delta(\eta^2+1) (\Delta \Psi)^2 \right),\label{DdimActionCSP}
\ee
where $\Delta \Psi \equiv (\partial_\eta.\partial_x + \rho)\Psi$. The only change from the helicity action \eqref{DdimAction} is the replacement
$\partial_\eta.\partial_x \rightarrow \partial_\eta.\partial_x+\rho$; so once again \eqref{DdimActionCSP} is invariant under gauge transformations 
\be
\delta_{\epsilon,\chi} \Psi \equiv (\eta.\partial_x - \frac{1}{2}(\eta^2+1) \Delta)\epsilon(\eta,x) + (\eta^2+1)^2 \chi(\eta,x) .\label{eq:GaugeTrans}
\ee

This section unpackages the physical content of this free action.  We first demonstrate that this action indeed propagates massless degrees of freedom of the ``continuous-spin'' type -- i.e. they transform faithfully under the Little Group, with a non-zero eigenvalue for the Pauli-Lubanski invariant.  To make the Little Group transformation properties more explicit, and to show that the action propagates precisely one CSP (i.e. one state for each helicity eigenvalue), we construct an explicit basis of gauge-inequivalent wavefunctions that solve the equations of motion following from \eqref{DdimActionCSP} and illustrate their transformation under the Little Group in \S\ref{Ssec:planewave}. In \S\ref{Ssec:tensor}, we recast this action and the resulting equation of motion in a tensor form similar to the Fronsdal action. 

\subsection{Mass and Spin}\label{Ssec:MassAndSpin}
To demonstrate that the action \eqref{DdimActionCSP} propagates continuous-spin particles, we show that the propagating degrees of freedom have eigenvalues $P^2 \Psi(\eta,x)= 0$ and $W^2\Psi(\eta,x) = -\rho^2\Psi(\eta,x)$ under the two invariants of the 3+1-dimensional Poincar\'e group, where $W^\mu = \frac{1}{2} \epsilon^{\mu\nu\rho\sigma} J_{\nu\rho} P_\sigma = \epsilon^{\mu\nu\rho\sigma} \eta_\nu {\partial_{\eta}}_\rho {\partial_x}_\sigma$ is the Pauli-Lubanski pseudo-vector.  This is to be contrasted with massless helicity particles, which have $P^2 =W^2 =0$, and with massive particles with spin, for which $P^2$ and $W^2$ are both non-zero.  Of course, in a gauge theory we should only demand these equalities up to pure gauge corrections:
\be
P^2 \Psi(\eta,x)= 0 + \rm{gauge} \quad \rm{and} \quad  W^2\Psi(\eta,x) = -\rho^2\Psi(\eta,x) +\rm{gauge}.
\ee

To show that the degrees of freedom are massless, we can simply reverse the argument of \S\ref{Ssec:bottomUpGauge}.  But because the $O(\rho^2)$ term in \eqref{DdimActionCSP} looks superficially like a mass term, it is worth going through the logic explicitly.  

Varying the action \eqref{DdimActionCSP} with respect to $\Psi$ yields the free equation of motion 
\be
\delta'(\eta^2+1) \Box \Psi - \frac{1}{2} \Delta\left( \delta(\eta^2+1) \Delta \Psi(\eta,x)\right)=0,\label{eomUnfixed}
\ee
which is gauge-invariant.  
As is often the case in gauge theories, the equation of motion suggests a choice of gauge in which the $\Psi$ equation of motion simplifies.    
In the gauge $\delta(\eta^2+1) \Delta \Psi = 0$, the e.o.m. simplifies to $\delta'(\eta^2+1) \Box \Psi = 0$. 
To see that this gauge can be reached, note that under $\epsilon$-type gauge transformations, 
$\delta_\epsilon \Delta \Psi(\eta,x) = -\frac{1}{2} (\eta^2+1)\Delta^2 \epsilon(\eta,x) + \Box \epsilon(\eta,x)$.  
The first term vanishes on the support of $\delta(\eta^2+1)$, while the second can be used to gauge away 
$\delta(\eta^2+1) \Delta \Psi$ using $\epsilon = - \frac{1}{\Box} \Delta \Psi$.
We call the gauge $\delta(\eta^2+1) \Delta \Psi = 0$ ``harmonic'' gauge because the equations of motion reduce to a massless wave-equation 
in this gauge. Writing $\Box \Psi = (\eta^2+1)^2 \beta(\eta,x)$ for any solution to this equation of motion, we can further gauge away $\beta$ by 
using a $\chi = - \frac{1}{\Box}\beta$ gauge transformation.
Thus, any solution to the equation of motion \eqref{eomUnfixed} is gauge-equivalent to a solution of
\be
\Box \Psi(\eta,x) = 0 \quad\mbox{and}\quad \delta(\eta^2+1) \Delta \Psi = 0. \label{harmonicGaugeEOM}
\ee
The first equation shows that the degrees of freedom propagated in $\Psi$ are massless.  
The harmonic gauge choice leaves a residual gauge 
freedom generated by arbitrary $\chi(\eta,x)$ and $\epsilon(\eta,x)$ with $\Box \epsilon (\eta,x)=\Box \chi (\eta,x)=0$.

Having shown that our action describes massless degrees of freedom, we can now study the action of the spin invariant $W^2$ on the field $\Psi$, working in harmonic gauge for simplicity.  
On the support of $\Box_x \Psi = 0$, the operator $W^2$ can be simplified to the form
\bea
W^2 \Psi &= &\left[ \eta^2 (\partial_\eta.\partial_x)^2 + (\eta.\partial_x)^2 \Box_\eta - 2 \eta.\partial_x \partial_\eta.\partial_x \eta.\partial_\eta \right] \Psi \\
& = & \left[ \eta^2 (\Delta - \rho)^2 + (\eta.\partial_x)^2 \Box_\eta - 2 \eta.\partial_x (\Delta - \rho) \eta.\partial_\eta \right] \Psi\\
& = & \left[ \eta^2 (\Delta - \rho)^2 + (\eta.\partial_x)^2 \Box_\eta - 2 \eta.\partial_x (\eta.\partial_\eta+1) (\Delta - \rho) \right] \Psi,
\eea
where in the last line we have used the commutator $[\Delta,\eta.\partial_\eta] = \Delta -\rho$.
Other useful commutation relations are 
\be
[\Delta,\eta^2] = 2 \eta.\partial_x,
\quad
[\eta.\partial_\eta,\eta^2] = 2 \eta^2,
\quad
[\partial_\eta^2,\eta^2] = 4 (\eta.\partial_\eta+2), \label{usefulCommutators}
\ee
of which the last is specific to 4 dimensions.  
For $\Psi$ in harmonic gauge, we can write 
\be
\Delta \Psi = (\eta^2+1) \alpha(\eta,x), \label{deDonderResidual}
\ee
where the residual $\alpha$ is unconstrained.  

Using this and the commutator relations \eqref{usefulCommutators} repeatedly, 
we obtain
\be
W^2  \Psi = -\rho^2 \Psi + \delta_\epsilon[(\eta.\partial_x \partial_\eta^2 -2 \partial_\eta.\partial_x \eta.\partial_\eta)\Psi - 2 \alpha] + \delta_\chi[\tfrac{1}{2}(\eta.\partial_x \partial_\eta^2 -2 \partial_\eta.\partial_x (\eta.\partial_\eta+2)) \alpha],
\ee
where $\alpha$ is the solution of \eqref{deDonderResidual}, and $\delta_\chi[\chi_T]$ denotes the gauge transformations obtained  from \eqref{eq:GaugeTrans} with $\chi(\eta,x) = \chi_T$ (and similarly for $\delta_\epsilon$).  We note that the pure-gauge residual is independent of $\rho$!

We thus see that $W^2 \Psi = -\rho^2\Psi$ up to gauge transformations, as claimed.  This shows that the action \eqref{DdimActionCSP} propagates CSP degrees of freedom with a definite spin-scale $\rho$ --- a considerable advance over \cite{CSP3}, which propagated a continuum of CSPs of \emph{every} real $\rho$.  
 We will see below, by constructing an explicit basis of states, that $\Psi$ represents precisely one CSP. 

\subsection{Plane Waves and Polarizations}\label{Ssec:planewave}
To identify the polarization content in $\Psi$ more explicitly, we now exhibit a convenient basis of solutions (up to gauge) to the harmonic-gauge equations of motion \eqref{harmonicGaugeEOM}.  We will decompose the solutions into helicity eigenmodes, as can always be done in  asymptoticly flat space. 
When the Pauli-Lubanski spin invariant $W^2$ vanishes, this decomposition is Lorentz invariant. 
When $W^2$ is nonzero, as we expect it is for $\rho\neq 0$, the helicity modes will mix under boosts. 
We will see this explicitly in the transformation of our basis wavefunctions.  

The helicity decomposition is readily carried out in a plane wave basis, $\Psi_{k}(\eta,x)=\psi(\eta,k)e^{-ik\cdot x}$,
where of course $k^2=0$ and the analytic function $\psi(\eta,k)$ satisfies
\be
\delta(\eta^2+1) (ik\cdot\partial_{\eta}+\rho) \psi(\eta,k) = 0.
\ee
As before, we can introduce a set of null frame vectors  
$q,\epsilon_{\pm}$ with $q\cdot k=1$, $\epsilon_{+}\cdot\epsilon_{-}=-2$, and all other products vanishing. 
The helicity operator is just
\be
R \Psi = \frac{1}{2}(\eta\cdot\epsilon_-\,\epsilon_+\cdot\partial_{\eta}-\eta\cdot\epsilon_+\,\epsilon_-\cdot\partial_{\eta})\Psi
\ee
For functions analytic in $\eta$, this operator necessarily has integer eigenvalues. 

One set of eigenstates satisfying the harmonic gauge condition --- which actually furnish a basis up to gauge transformations --- is
\be
\psi_{+h}(\eta,k) = (\eta\cdot \epsilon_+)^{h}e^{i\rho\eta\cdot q} \quad \psi_{-h}(\eta,k) = (\eta\cdot \epsilon_-)^{|h|}e^{i\rho\eta\cdot q}.\label{eq:basis}
\ee

To see that these are a basis, we first consider an arbitrary function $\psi(\eta.k,\,\eta.q,\,\eta.\epsilon_+,\,\eta.\epsilon_-)$ at null momentum $k$ (we leave the $k$-dependence of $\psi$ implicit).  The harmonic gauge condition implies $(i k.\partial_\eta+\rho)\psi(\eta) = (\eta^2+1) \alpha(\eta)$; when $k^2 =0$, transforming by a gauge parameter $\epsilon$ satisfying $(i k.\partial_\eta+\rho)^2 \epsilon = 2 \alpha$ brings $\psi$ into a gauge where $(i k.\partial_\eta + \rho)\psi = 0$ everywhere (not just on the $\eta^2+1=0$ surface), so that we can write $\psi = e^{i\rho \eta.q} \left[ f(\eta.\epsilon_+,\, \eta.\epsilon_-) + \eta.k \; g(\eta.k,\, \eta.\epsilon_+,\, \eta.\epsilon_-)\right]$, where we have explicitly separated out the $\eta.k$-independent part of $\psi$.  Transforming by $\epsilon = - e^{i\rho \eta.q} g$ removes all $\eta.k$-dependence, so that $q.\partial_\eta \psi = 0$.  At this point, the remaining gauge freedom that preserves the conditions $(ik.\partial_\eta +\rho)\psi  = q.\partial_\eta \psi = 0$ are those generated by $\epsilon = e^{i\rho \eta.q} \,q.\eta \,h(\eta.\epsilon_+,\, \eta.\epsilon_-)$, which generate $\delta \psi = \frac{1}{2}e^{i\rho \eta.q} (\eta.\epsilon_+ \eta.\epsilon_- - 1)\, h(\eta.\epsilon_+,\,\eta.\epsilon_-)$.  These can be used to remove any dependence of $\psi$ on the product $\eta.\epsilon_+\,\eta.\epsilon_-$, so that all that remains is a sum of the basis functions \eqref{eq:basis}, $\psi(\eta) = e^{i\rho \eta.q} (f_+(\eta.\epsilon_+) + f_-(\eta.\epsilon_-))$.  
Thus, any function satisfying the harmonic gauge conditions and equation of motion can be decomposed 
into a linear combination of basis functions of the form \eqref{eq:basis} plus pure gauge, and we have already argued that a general solution to the covariant equation of motion \eqref{eomUnfixed} can be brought into harmonic gauge.  
In fact, the argument above applies whether or not $q$ is null, so we could for example reach a ``Coulomb-like'' gauge where $q^\mu =(1/k^0,0,0,0)$ and $\nabla_\eta.\nabla_x \psi = 0$.  
This construction also justifies the description of $\psi$ as a sum of holomorphic and anti-holomorphic functions on a complex plane introduced in \S\ref{Ssec:bottomUpGauge}.

The basis functions \eqref{eq:basis} are also orthonormal under the natural inner product 
\be
\braket{\psi_h(k)}{\psi_{h'}(k)} \equiv \int d^4\eta \delta'(\eta^2+1) \psi_h(\eta,k)^* \psi_{h'}(\eta,k) = \delta_{h,h'}.
\ee
Evaluating this integral using the Euclidean $\eta$-space techniques is very simple, and is done explicitly in Appendix \ref{Sapp:orthonormality}.

Given an arbitrary wave-function $\Psi(\eta,k)$ in momentum space satisfying the equation of motion, we
can project out helicity components $a^*_h(k)$ (for positive energy null $k^0>0$) as
\be
\frac{a^*_h(k)}{2|k|} =  \int d^4\eta \delta'(\eta^2+1) \psi_h(\eta,k) \Psi(\eta,-k)\label{ahdef}
\ee
where the normalization is chosen to coincide with relativistic conventions. 
This projection is gauge invariant. The variation of \eqref{ahdef} under a gauge transformation is
\bea
\int d^4\eta \delta'(\eta^2+1) \psi_h(\eta,k)\left( -ik \cdot \eta - \tfrac{1}{2} (\eta^2+1) \left(-ik \cdot \partial_\eta+\rho\right)\right) \epsilon && \nonumber \\
= \int d^4\eta \psi_h(\eta,k)\left(-ik \cdot \partial_\eta+\rho\right)\left(\delta(\eta^2+1)\epsilon\right) &&
\eea
which vanishes after integrating by parts in $\eta$ because $\delta(\eta^2+1)\left(ik \cdot \partial_\eta+\rho\right) \psi_h(\eta,k)=0$ (i.e. $\psi_h$ satisfies the harmonic-gauge condition).

The helicity decomposition for a real field $\Psi(\eta,x)$ is therefore
\be
\Psi(\eta,x) = \int \frac{d^3k}{2|k|}\sum_h\left( a_h(k)\psi_h(\eta,k)e^{-ik\cdot x}+a^*_h(k)\psi_h^*(\eta,k)e^{ik\cdot x}\right). \label{eq:modeexpansion}
\ee
where $a^*_h(k)$ denotes complex conjugation and we are allowed of course to add a pure gauge term to the right hand side. 

Now we can see how the gauge symmetry allows CSPs to be encoded in the wavefunction
$\Psi$. 
Let us consider how $a^*_h(k)$ defined by \eqref{ahdef} transforms under the combinations 
of rotations and boosts that leave $k$ invariant, namely transformations generated by $T_{\pm}=\epsilon_{\pm}\cdot W$. 
These generators should act like raising and lowering operators on $a^*_h(k)$.
Starting from the covariant transformation rule $U(\Lambda)^{\dagger}\Psi(\eta,k)U(\Lambda)\equiv \Psi(\Lambda^{-1}\eta,\Lambda^{-1}k)$,
we have
\bea
U(\Lambda)^{\dagger}\frac{a_h^*(k)}{2|k|}U(\Lambda) &\equiv& \int d^4\eta \delta'(\eta^2+1) \psi_h(\eta,k) \Psi(\Lambda^{-1}\eta,-\Lambda^{-1}k) \nonumber \\
&=& \int d^4\eta \delta'(\eta^2+1) \psi_h(\Lambda \eta,k) \Psi(\eta,-\Lambda^{-1}k) \nonumber
\eea
Specializing to transformations $U(\Lambda)=e^{i(\beta_+T_++\beta_-T_-)}$ generated by $T_{\pm}$, and working infinitesimally,
we have
\be
[T_{\pm},a^*_h(k)] = \int d^4\eta \delta'(\eta^2+1)[T_{\pm},\psi_h(\eta,k)]\Psi(\eta,-k) \label{eq:LGonMode},
\ee
where we have used the fact that $T_{\pm}$ annihilates $k$. We are free to work in the harmonic gauge $\delta(\eta^2+1)\Delta\Psi=0$, where any pure-gauge terms in $[T_{\pm},\psi_h(\eta,k)]$
 give vanishing contributions to the integral \eqref{eq:LGonMode}, by an integration-by-parts argument analogous to that used in deriving \eqref{eq:modeexpansion}. 
For $h>0$, we have  $\psi_h(\eta,k) = (\eta\cdot \epsilon_+)^{h}e^{i\rho\eta\cdot q}$ and 
\bea
[T_{+},\psi_h(\eta,k)] &=& i(\eta\cdot k\,\epsilon_{+}\cdot\partial_{\eta}-\eta\cdot\epsilon_{+}\,k\cdot\partial_{\eta})\psi_h(\eta,k) \nonumber \\
&=& \rho \eta\cdot\epsilon_+\psi_h(\eta,k) = \rho \psi_{h+1}(\eta,k), \nonumber
\eea
so that $T_+$ acts as expected on the mode $a^*_h(k)$, i.e. $[T_+,a^*_h(k)] = \rho a^*_{h+1}(k)$.
On the other hand
\bea
[T_{-},\psi_h(\eta,k)] &=& i(\eta\cdot k\,\epsilon_{-}\cdot\partial_{\eta}-\eta\cdot\epsilon_{-}\,k\cdot\partial_{\eta})\psi_h(\eta,k) \nonumber \\
& = & -2i\,h\, \eta.k \, \psi_{h-1}(\eta,x) + \rho\, \eta\cdot\epsilon_{-}\psi_{h} \\
& = & \rho\, \psi_{h-1} + \delta_{\epsilon}[- 2(h + i \eta\cdot q) \psi_{h-1} ].
\eea
To see that the last two lines are indeed equivalent up to pure gauge terms (which contribute nothing to the integral \eqref{eq:LGonMode}), it is useful to note that 
$g^{\mu\nu}=-\frac{1}{2}\epsilon^{(\mu}_{+}\epsilon^{\nu)}_{-} + q^{(\mu}k^{\nu)}$ so that $\eta^2 = - \eta.\epsilon_+ \,\eta.\epsilon_-+2 \eta.q \,\eta.k$ and therefore $\epsilon = -2i (q\cdot \eta) \psi_{h-1}$ generates the gauge transformation
\bea
\delta \psi = (i k\cdot \eta - \tfrac{1}{2}(\eta^2+1)(i k\cdot\partial_{\eta}+\rho))\epsilon &=&  2 k\cdot\eta \, q\cdot\eta \, \psi_{h-1}-(\eta^2+1)\psi_{h-1} \nonumber \\
&=& (2k\cdot\eta \,q\cdot\eta -\eta^2)\psi_{h-1} - \psi_{h-1} \nonumber \\
&=& \epsilon_+\cdot\eta\,\epsilon_-\cdot\eta \psi_{h-1} - \psi_{h-1}, \nonumber
\eea
while $\epsilon = -2 h \psi_{h-1}$ simply generates $\delta \psi = -2i\, h \eta.k \psi_{h-1}$.
We therefore obtain 
\be
[T_{-},\psi_h(\eta,k)] \simeq \rho \psi_{h-1}, 
\ee
so that going back to \eqref{eq:LGonMode} we find
\be
[T_{\pm},a^*_h(k)] = \rho a^*_{h\pm 1}(k) \label{eq:LGonModes}
\ee
as expected for the polarizations of a CSP with a spin scale $W^2=-\rho^2$. Similar manipulations show that $T_{\pm}$
act correctly for $h\leq 0$ as well. 

We will take up a covariant quantization of this theory in \cite{CSPjenny}, but much of the free quantum theory is already easy to 
anticipate from the mode expansion. The coefficients $a_h(k)$ will become annihilation operators with commutation relations
\be
[a_h(k),a_{h'}(k')^*]=2|k|\delta^3(k-k')\delta_{hh'}, \label{eq:CommRelations}
\ee
with the interpretation of creating or annihilating states with momentum $k$ and polarization $h$.
This fact, together with \eqref{eq:LGonModes}, shows that the quantum theory indeed describes a single CSP. 

\subsection{The Action and Equation of Motion in Tensor Form}\label{Ssec:tensor}
To gain more intuition, it is useful to decompose the action into tensor fields as we did for $\rho=0$ in  Section \ref{Ssec:tensorDecomposition}. As before, we decompose $\Psi(\eta,x)$ into tensor fields as 
\bea
\Psi(\eta,x) 
& = & \sum_{n}  c_n \left( \eta^n - \tfrac{1}{2} g\circ \eta^{n-2} (\eta^2+1) \right) \phi^{(n)},\label{FronsdalDecomposition2}
\eea
with the coefficients $c_n = 2^{n/2}$ chosen so that the fields $\phi^{(n)}$ are canonically normalized.
We can again fix a gauge where the $\phi^{(n)}$ are double-traceless using $\chi$ gauge-invariance; the gauge transformations that preserve this double-traceless gauge are parametrized by 
$\epsilon(\eta,x) = \sum_{n} (n+1) c_{n+1} \eta^{n} \epsilon^{(n)}(x)$ where $\epsilon^{(n)}(x)$ is an arbitrary \emph{traceless} rank-$n$ tensor.  
The component fields vary under these gauge transformations as
\be
\delta \phi^{(n)} = \partial \circ \epsilon^{(n-1)} + \frac{\rho}{\sqrt{2}}  \left[ \epsilon^{(n)} + \tfrac{1}{n(n-1)} g \circ \epsilon^{(n-2)} \right].
\ee

The action for the Fronsdal-like fields can be found by the same methods used in \S\ref{Ssec:tensorDecomposition}.  Here
\be
\Delta \Psi \simeq \sum_{n\ge 1} \eta^{n-1}\cdot \left(n c_n (\partial_x\cdot \phi^{(n)} - \tfrac{1}{2}\partial_x \cdot {\phi^{(n)}}')+\rho c_{n-1} \phi^{(n-1)}\right),
\ee
where ${\phi^{(n)}}' = Tr \phi^{(n)}$ and $\simeq$ denotes that terms proportional to $\eta^2+1$, which vanish on the $\delta$-function support, have been dropped.
With the inclusion of the $O(\rho)$ term this is no longer traceless (but it is double-traceless), so when we compute $G(i\partial_\eta) (\Delta \Psi)^2$ as in \S\ref{Ssec:tensorDecomposition} using the series expansion for the generating function $G(i\partial_\eta) = J_0(\sqrt{\partial_\eta^2})$ we must keep terms where zero or one $\partial_\eta^2$ act on the series expansion of each $\Delta \Psi$.  The resulting action can be written as 
\be
S=\sum_{n} \int d^4 x \tfrac{1}{2} (\partial_\mu \phi^{(n)})^2 - \tfrac{n(n-1)}{8} (\partial_\mu {\phi^{(n)}}')^2 - \frac{n}{4} (D_\rho^{(n-1)})^2
\ee
where
\be
D_\rho^{(n-1)}(x) \equiv \partial_x\cdot \phi^{(n)} - \tfrac{1}{2}\partial_x \circ {\phi^{(n)}}'+  \frac{\rho}{\sqrt{2} n} \left(\phi^{(n-1)}- \tfrac{1}{2 (n-1)} g\circ {\phi^{(n-1)}}' \right) - \frac{\rho}{2 \sqrt{2}}{\phi^{(n+1)}}'.
\ee
The action, which was rank-diagonal at $O(\rho^0)$, exhibits rank-mixing at $O(\rho)$.  Of course, the action becomes rank-diagonal in harmonic gauge (which corresponds to $D_{\rho}^{(n-1)} = 0$), but then the gauge condition itself mixes ranks.  

Similarly, the CSP equation of motion mixes tensors of different ranks: 
\be
- \Box \phi^{(n)} + \partial \circ D_{\rho}^{(n-1)} + \tfrac{\rho}{\sqrt{2}} D_{\rho}^{(n)} + \tfrac{\rho}{\sqrt{2} n(n-1)} g \circ D_{\rho}^{(n-2)} = 0.\label{CSPtensoreom}
\ee
While the tensor decomposition makes manifest that we smoothly recover the Schwinger-Fronsdal equations and action in the $\rho\rightarrow 0$ limit,
it is more cumbersome for non-zero $\rho$ than the simple vector superspace description. 
For one thing, the equation of motion \eqref{CSPtensoreom} is a bit complicated (of course, it may be possible that a suitable field redefinition simplifies it --- in particular, it would be interesting to see whether the equation of motion previously obtained by \cite{BekaertMourad} is equivalent to this one).  
Another drawback --- and a much more invariant one --- is that our helicity eigenmodes 
$\psi(\eta,k) = (\eta.\epsilon_\pm)^{|h|} e^{i\rho \eta.q}$ (for $q$ with $q\cdot k=1$ and $q\cdot\epsilon_\pm = 0$) from \S\ref{Ssec:planewave} do not correspond to a single tensor mode when $\rho \neq 0$.  For example, for $h=0$ we have $\phi^{(0)} = 0$, $\phi^{(1)}_{\mu} = i \rho q_\mu /\sqrt{2}$, $\phi^{(2)}_{\mu\nu} = - \rho^2 q_\mu q_\nu/2$, etc.  It is tempting to look for a Lorentz-covariant field redefinition where the helicity $\pm h$ modes are encoded only in a rank-$h$ tensor, but in fact such a representation is at odds with the action of the little group on single-particle CSP states.  In particular, any rank-$h$ tensor is annihilated by $(T_{\pm})^{h+1}$, while this would simply raise (lower) the helicity $h$ mode of a CSP to a mode with helicity $(2h+1)$ or $-1$.  Thus, a covariant representation of \emph{any} CSP state in terms of tensors must involve infinitely many non-zero tensor components, so that $T_{\pm}$ can act arbitrarily many times without annihilating the state. 
These and other considerations underscore the value of the vector superspace formulation emphasized in this paper.

\section{A Window on CSP Interactions}\label{Sec:physical}

Though we have thus far focused on the free action, it is straightforward to couple the gauge field $\Psi$ to a background current by adding a source term  
\be
S_{int} = \int d^4 x [d^4 \eta]  \delta'(\eta^2+1) \Psi(\eta,x) J(\eta,x). \label{eq:source}
\ee
This is, in fact, the most general linear current that we can couple to the degrees of freedom of $\Psi$ in the first 
neighborhood of the $\eta^2+1=0$ surface. 
The coupling is allowed by gauge-invariance so long as $J$ satisfies a ``continuity condition'' 
\be
\delta(\eta^2+1) (\partial_x.\partial_\eta + \rho) J = 0,
\ee
or in other words
\be
(\partial_x.\partial_\eta + \rho) J = (\eta^2+1)\alpha(\eta,x), \label{eq:continuity}
\ee
for arbitrary analytic $\alpha(\eta,x)$.   

This section explores some simple features of this current coupling, for $\rho=0$ and non-zero $\rho$.  In \S\ref{Ssec:helicityInteractions} we sketch how standard scattering amplitudes involving helicities 0, 1, and 2 can be obtained in the vector superspace formalism.  \S\ref{Ssec:rhoDeformedContinuity} describes how the usual conservation conditions are modified with non-zero $\rho$, and the pattern of non-zero currents at all ranks suggested by the helicity correspondence of CSP-emission amplitudes.  Finally, \S\ref{Ssec:softEmission} demonstrates the consistency of this formalism with the candidate single-CSP amplitudes found in \cite{CSP1,CSP2}, albeit with an ad hoc and non-local ansatz for current matrix-elements.

\subsection{Helicity Interactions in Vector Superspace}\label{Ssec:helicityInteractions}
It is straightforward to calculate simple helicity amplitudes using $\eta$-space, and in fact this formalism is appealingly universal, regardless of whether the emitted particle has helicity 0, 1, or 2.  Before doing these calculations, however, we pause to make contact between the vector superspace continuity condition \eqref{eq:continuity} and the more familiar conservation conditions, and in particular to comment on the significance of the arbitrary function $\alpha$.

It will be convenient to expand $J(\eta,x)$ into tensor components so that the current-coupling in tensor components is rank-diagonal, i.e. so that \eqref{eq:source} reduces to ${\sum_n (-1)^{n} \phi^{(n)}(x)\cdot J^{(n)}(x).}$  This is achieved by decomposing $J(\eta,x)$ as
\be
J(\eta,x) = \sum_n 2^{n/2} \left( \eta^n + \frac{1}{2(n-1) }\eta^{(n-2)} \circ g\right) J^{(n)} = \sum_n 2^{n/2} \eta^n \left(J^{(n)} + \frac{n+2}{2} {J^{(n+2)}}' \right).
\ee
In this case, allowing non-zero $\alpha$ is analogous (and, for $\rho=0$, equivalent) to the well-known consistency of coupling high-spin fields to currents that are only conserved up to traces (i.e. ``weakly conserved''), e.g. $\partial^\mu J^{(3)}_{\mu\nu\rho} (x) = g_{\nu\rho} \alpha(x)$ with arbitrary $\alpha$.  Thus, the possibility of non-zero $\alpha$ can be ignored for helicities $\le 2$.  For $\rho = 0$, the continuity condition does not constrain $J^{(0)}$ and imposes the usual conservation requirements $\partial.J^{(1)} = 0$, and $\partial.J^{(2)}$ for the rank-1 and 2 currents.

Returning to the vector superspace formalism, it is simple to calculate emission amplitudes for a polarization $h$ component of $\Psi$ with momentum $k$ in the presence of the background $J(\eta,x)$ which we assume to have only $J^{(0)}$, $J^{(1)}$, and/or $J^{(2)}$ non-zero, i.e $J(\eta,x) = J^{(0)}+ \sqrt{2} \eta\cdot J^{(1)} + (2 \eta^2\cdot J^{(2)} +{J^{(2)}}') .$
Representing the single particle states as $a_h^{\dagger}(k)\ket{0}$, the amplitude is just
\bea
{\mathcal A}_J(\{k,h\}) &=& \int d^4x d^4\eta \delta'(\eta^2+1)\bra{k,h}\Psi(\eta,x)\ket{0}J(\eta,x) \nonumber \\
&=& \int d^4x d^4\eta \delta'(\eta^2+1) \psi_h^*(\eta,k)e^{ik\cdot x}J(\eta,x) \nonumber \\
&=&  \int d^4\eta \delta'(\eta^2+1) \psi_h^*(\eta,k)\tilde{J}(\eta,-k)
\eea
where $\tilde{J}(\eta,-k)$ is the Fourier transform of $J(\eta,x)$. 
For the current $J(\eta,x)$ given above, we find that all helicity modes with $|h|>2$ have vanishing emission 
amplitude, while for $|h| \leq 2$ we obtain
\bea
{\mathcal A}_J(\{k,h=0\}) &=& \tilde{J_0}(-k)\\
{\mathcal A}_J(\{k,h=1\}) &=& \epsilon_{+,\mu}(k)\tilde{J_1}^{\mu}(-k)\\
{\mathcal A}_J(\{k,h=2\}) &=& \epsilon_{+,\mu}(k)\epsilon_{+,\nu}(k)\tilde{J_2}^{\mu\nu}(-k)
\eea
which precisely reproduces familiar results for helicity degrees of freedom. 

We can similarly calculate current-current correlators, and hence matter amplitudes provided the matter sector
can furnish an appropriately conserved current (which we know how to do for $\rho=0$ of course). 
As will be discussed in \cite{CSPjenny}, the harmonic gauge 
propagator for $\Psi$ is
\be
 \delta'(\eta^2+\alpha) \delta'(\overline{\eta}^2+\alpha)\langle 0| T\Psi(\eta,x)\Psi(\overline{\eta},y)|0\rangle = \delta'(\eta^2+\alpha)\delta^{(4)}(\eta-\overline{\eta})D_F(x-y),
\ee
where $D_F(x-y)$ is the Feynman propagator. Suppose we have a particle of type A with current $J_A(\eta,x)$
and of type B with current $J_B(\eta,x)$. Then the first order scattering amplitude $A(A,B\rightarrow A'B')$
mediated by $\Psi$-exchange is
\bea
{\mathcal A}(A,B\rightarrow A'B') &=& \int d^4x d^4\eta \delta'(\eta^2+1)\int d^4y d^4\eta' \delta'(\eta'^2+1)\times \nonumber \\
&&\bra{A'}J_A(\eta,x)\ket{A}\bra{0}T\Psi(\eta,x)\Psi(\eta',y)\ket{0}\bra{B'}J_B(\eta',y)\ket{B} \nonumber \\
&=& \int d^4xd^4y d^4\eta \delta'(\eta^2+1)\bra{A'}J_A(\eta,x)\ket{A}D_F(x-y)\bra{B'}J_B(\eta,y)\ket{B} \nonumber
\eea
For a scalar current $J(\eta,x)=J_0(x)$, this obviously reproduces standard results. 
For a vector current $J(\eta,x)=\eta\cdot J_1(x)$, the $\eta$-integration yields a $g^{\mu\nu}$
contraction between $J_1^{\mu}(x)$ and $J_1^{\nu}(y)$, again as expected.
For a tensor current, the $\eta$-integration yields the expected symmetric tensor contractions.
In fact, it is straightforward to re-cast simple theories like QED  in this language, 
and calculations of correlators and scattering amplitudes are straightforward, as the above examples illustrate. 
It would be interesting to express the self-interactions of Yang-Mills theories and General Relativity in vector superspace language,
but we have not done so.  

\subsection{Deformed Conservation Conditions for Currents Coupled to CSPs}\label{Ssec:rhoDeformedContinuity}
For $\rho=0$, we could consistently consider sources with only one non-zero $J^{(n)}$ at a time.  
Of course, in flat space non-trivial conserved currents only exist up to rank-2. 
For $\rho\neq 0$, the continuity condition \eqref{eq:continuity} in general requires that if any single component of $J(\eta,x)$ is non-zero, all higher rank components must be non-zero as well.  This is easy to see by breaking down \eqref{eq:continuity} into components: 
\bea
\partial \cdot J_1 + \rho J_0 &=& 0 \nonumber \\
\partial \cdot J_2^{\mu} + \rho J_1^{\mu} &=& 0 \nonumber \\
\tsub{\partial \cdot J_3^{\mu\nu} + \rho J_2^{\mu\nu}} &=& 0 \nonumber \\
&...& \nonumber
\eea
where $\tsub{\dots}$ denotes the traceless part of the enclosed tensor.  If the scalar current $J_0(x)$ is non-zero, the first continuity condition 
requires a non-zero $J_1^{\mu}(x)$ at $O(\rho)$, with non-zero divergence. The next continuity condition in turn implies a non-zero $J_2^{\mu\nu}(x)$ at $O(\rho^2)$, and so on.
If we instead started with a {\it conserved} vector current $J_1^{\mu}(x)$ at $O(\rho^0)$, the continuity 
condition would imply that there is a non-zero $J_2^{\mu\nu}(x)$ at $O(\rho)$ with non-trivial divergence, though $J_0$ can consistently vanish. 
Finally, a similar pattern of currents can be initiated by a {\it conserved} rank-2 tensor, with vanishing scalar and vector current, and all 
higher ranks of successively higher order in $\rho$.  
This structure is a field theory version of the {\it helicity correspondence} discovered in \cite{CSP2} 
-- all tree level amplitudes that consistently factorize and obey perturbative unitarity constraints all approach 
helicity-0,1, or 2 amplitudes in the high energy or non-relativistic limit (or $\rho\rightarrow 0$ limit).

\subsection{Making Contact with Soft-Emission Amplitudes}\label{Ssec:softEmission}

We do not yet know of any {\it local} matter sector that furnishes an appropriately conserved current that can serve as a source for \eqref{eq:source} when $\rho\neq 0$.
{\bf This is the central open problem to solve in formulating a complete CSP-theory.} 
If we sacrifice manifest locality, then we can readily guess appropriate current matrix elements, and they will in general (as required by Lorentz symmetry) reproduce the soft factors found in \cite{CSP1,CSP2}.  
As a specific example, we consider an ansatz for a correlation function $\bra{p'}J(\eta,x)\ket{p^*}$.
Here, the notation $\ket{p^*}$ means that instead of a matrix element with an external state, 
we are instead considering the correlator of a matter field at a (slightly) off-shell momentum $p^*$ with the current. 
We then expect that the soft factor for emission of a soft CSP with momentum $k$ and polarization $h$ is of the form
\be
S(\{k,h \},p) = \int d^4x d^4\eta \delta'(\eta^2+1) \bra{k,h}\Psi(\eta,x)\ket{0}\bra{p'}J(\eta,x)\ket{p^*}
\ee
where $k+p'=p$. One guess that satisfies the continuity condition \eqref{eq:continuity} is 
\be
\bra{p'}J(\eta,x)\ket{p^*} = \int d^4 q e^{iq\cdot x}e^{i\rho\frac{\eta\cdot (p+p')}{k\cdot (p+p')}} 
\ee
This guess is one of many that satisfy \eqref{eq:continuity}, and actually satisfies a stronger continuity condition than is required, since $\Delta J =0$ vanishes even off the $\delta$-function support (i.e. it has $\alpha = 0$)\footnote{However, this guess cannot be directly applied to matter-matter scattering mediated by a CSP.}. Integrating, we obtain (for $h\geq0$)
\be
S(\{k,h \},p) = \int d^4\eta \delta'(\eta^2+1) (\eta\cdot \epsilon_+(k))^h e^{i\rho\frac{\eta\cdot p}{k\cdot p}} \nonumber
\ee
These integrals are easy to do using the generating function formula, or by direct integration in euclidean $\eta$-space
as is illustrated by example in Appendix \ref{Sapp:orthonormality}. Using standard integral representations of Bessel functions,
we have
\be
S(\{k,h \},p) = e^{i h \arg z} J_h(\rho z),
\ee
where $z\equiv \frac{\epsilon_{+}\cdot p}{k\cdot p}$. This is precisely the soft factor found in \cite{CSP1,CSP2}! The underlying gauge symmetry of our field 
theory formalism is maintained by the ansatz above, so this answer was essentially guaranteed by Lorentz symmetry, 
but it is reassuring to see that such CSP soft factors can be obtained from our field theory. 

\section{Comments on Gauge-Invariant Operators}\label{sec:GIcomments}

While the preceding discussion focused on the consistency requirements for a CSP to couple to a background current, we turn now to explore physically significant quantities that can be constructed from the CSP field $\Psi(\eta,x)$ itself.  Our focus in \S\ref{Ssec:energyMomentum} is on the stress-energy tensor, which is not gauge-invariant, and its spatial integrals, which \emph{are} gauge-invariant.  This situation is quite reminiscent of general relativity, where the energy density of a gravitational wave at a point is not physically observable, but the average energy density in a sufficiently large physical region is\cite{MTW}.  In \S\ref{sec:GIoperators} we turn to single-field gauge-invariant operators.  It is straightforward to show that no \emph{local} gauge-invariant operators exist that involve two or fewer derivatives, except for those which are proportional to the free equation of motion and therefore vanish on radiation. 
This is again reminiscent of GR, albeit slightly \emph{less} local (the Riemann curvature $R^{\mu\nu\rho\sigma}$ is not invariant under a a full coordinate transformation, but it \emph{is} invariant under the linearized gauge transformations).

But then, in light of the helicity correspondence, what happens to operators like $F^{\mu\nu}(x)$?  We will show that there are operators local in time, but non-covariant and spatially non-local, that approach these local field-strengths in the $\rho \rightarrow 0$ limit.  In fact, with a covariant quantization one can show that these objects commute at arbitrary spacelike separation, and are therefore causal.  It is still not clear that these are physically relevant objects, but they do indicate that Cauchy problems involving only physical data continue to be well-posed. 

\subsection{Energy, Momentum, and Angular Momentum of a CSP}\label{Ssec:energyMomentum}

Canonical \Poincare generators can be obtained in the usual way from the Belinfante stress-energy tensor
\be
\Theta^{\mu\nu}(x) \equiv -g^{\mu\nu}{\cal L}(x) + \int d^4\eta \frac{\delta {\cal L}}{\delta (\partial_\mu \psi)} \left(\partial^{\nu} \psi\right) 
+ \partial_{\kappa} \left(A^{\kappa \mu\nu} - A^{\mu\kappa\nu}-A^{\nu\kappa\mu} \right)
\ee
where 
\be
A^{\kappa\mu\nu} \equiv \frac{i}{2} \int d^4 \eta \frac{\delta{ \cal L}}{\delta (\partial_{\kappa} \psi)} {\cal J}^{\mu\nu} \psi(\eta,x),
\ee
and ${\cal J}^{\mu\nu} \equiv i \eta^{[\mu}\partial_{\eta}^{\nu]}$ are the Lorentz generators on the $\eta$-space.
Here and throughout this section, we use $\eta^{[\mu}\partial_{\eta}^{\nu]} = \eta^{\mu}\partial_{\eta}^{\nu} - \eta^{\nu}\partial_{\eta}^{\mu}$. 
Because $\Theta^{\mu\nu}$ is conserved and symmetric, the three-tensor 
\be
{\cal M}^{\kappa\mu\nu} \equiv x^{\mu} \Theta^{\kappa\nu} - x^\nu\Theta^{\kappa\mu} 
\ee
is also conserved.  Its conserved charges
\bea
J^{\mu\nu} & \equiv & \int d^3 x {\cal M}^{0\mu\nu} = \int d^3x  \left( x^{[\mu} T^{0\nu]} - 2 A^{0\mu\nu} \right)\\
& = & -\int d^3 x \; x^{[\mu} g^{0\nu]}{\cal L}(x) + \int d^3x d^4\eta \frac{\delta {\cal L}}{\delta \dot \psi} 
\left( x^{[\mu}  \partial^{\nu]} + \eta^{[\mu}\partial_{\eta}^{\nu]}\right) \psi(\eta,x)
\eea
 of course generate homogeneous Lorentz transformations on fields.  
The total energy, momentum, and angular momentum in harmonic gauge are
\bea
H &=& \int d^3x  \Theta^{00} = \int d^3x [d^4\eta] \delta'(\eta^2+1)\left(\frac{1}{2}\dot \psi^2 + \frac{1}{2}|\nabla\psi|^2 \right) \\
P^{i} &=& \int d^3x  \Theta^{0i} = \int d^3x [d^4\eta] \delta'(\eta^2+1) \dot \psi \partial^i  \psi \\ 
J^{ij}  &=&  \int d^3x  \Theta^{ij} = - \int d^3x [d^4\eta] \delta'(\eta^2+1) \dot \psi \left( x^{[i} \partial^{j]} + \eta^{[i} \partial_\eta^{j]} \right) \psi.
\eea
The Belinfante tensor $\Theta^{\mu\nu}$ and the associated ${\cal M}^{\mu\nu\rho}$ change under gauge transformations (even those that maintain harmonic gauge), but only by total derivatives so that these spatial integrals are gauge-invariant.  For example, under gauge transformations that maintain harmonic gauge $$\delta {\cal M}^{0ij}(x) = \int [d^4\eta] \dot \psi(\eta,x) R^{ij} \Delta(\delta(\eta^2+1) \epsilon(\eta,x)) + \Delta(\delta(\eta^2+1) \dot\epsilon(\eta,x)) R^{ij} \psi(\eta,x),$$ where $R^{ij} = x^{[i} \partial_x^{j]} +\eta^{[i} \partial_\eta^{j]}$.  Because $\delta(\eta^2+1) \Delta \psi = 0$ in harmonic gauge, this can be rewritten as a total derivative $-{\partial_x}_\mu \int [d^4\eta] \delta(\eta^2+1)  \left( ({\partial_\eta}^\mu \dot \psi(\eta,x)) R^{ij} \epsilon(\eta,x) + \dot\epsilon(\eta,x)) \partial_{\eta}^\mu R^{ij} \psi(\eta,x)\right)$.  As usual, although the total time-derivative term cannot be removed by integration by parts in the definition of $J^{ij}$, it can be rewritten in terms of total spatial 
derivatives using the equation of motion $\Box \psi = 0$ and the restriction $\Box \epsilon = 0$  on the gauge parameter.

The non-covariance of the Belinfante stress-energy tensor  is not surprising.  Even for $\rho=0$, the same phenomenon is observed in linearized general relativity, where gravitational waves do not have a well-defined stress-energy density.  The absence of a conserved, gauge-invariant stress-energy density for CSPs does motivate the expectation that, like in GR, the flat coordinate space $x^{\mu}$ used in our free formulation will have no invariant meaning in the presence of CSP interactions.  

In GR, gravitational waves do of course carry energy-momentum, and if we integrate $T^{\mu\nu}$ over a spacetime region of size $L$ much larger than the wavelength $\lambda$ of gravitational radiation, then the spatial average over this region of $T^{\mu\nu}$ is gauge-invariant up to effects suppressed by $\lambda/L$.  This follows simply from the fact that $T^{\mu\nu}$ is gauge-invariant up to spatial total derivatives (after using the equation of motion to remove time total-derivatives).  Because the same is true for the radiation in our CSP theory, there is no obstacle to answering the physical question of how much energy a CSP wave carries through a large spatial region, at least at the free level.  

We can also ask if the spatially averaged stress energy tensor $T_{avg}^{\mu\nu}(x)$ has properties compatible 
with causality in the quantum theory. 
In the quantum theory, it follows from \eqref{eq:CommRelations} that the gauge potentials $\Psi(\eta,x)$ (in a Feynman-like gauge) 
satisfy familiar commutation relations  
\be
\delta'(\eta^2+1) \delta'(\eta'^2+1) [\Phi(\eta,x), \Phi(\eta',x')] = \delta'(\eta^2+1) \delta^{(4)}(\eta'-\eta) \Delta(x-y), \label{eq:commutator}
\ee
where $\Delta(x-y) = \int \frac{d^3p}{2|p|} [ e^{-i p.(x-y)} - c.c.]$ is the usual scalar-field commutator.  It follows that in this gauge $[T^{\mu\nu}(x), T^{\rho\sigma}(x')] = 0$ for $x-x'$ spacelike, and therefore that the spatial averages commute in any gauge, provided the averaging regions are small compared to $|x-x'|$.

\subsection{Some Single Field Gauge Invariant Operators}\label{sec:GIoperators}
Besides the stress-energy carried by a wave, another very important class of physical objects in familiar gauge theories are the field-strengths.  Of course, in non-Abelian gauge theories and in GR these are not gauge invariant beyond linear level.  But $F^{\mu\nu} = \partial^{[\mu} A^{\nu]}$ is invariant under the free gauge transformation of a vector field, and so is the linearized curvature for a helicity-2 field, $R^{\mu_1\mu_2\nu_1\nu_2} = \partial^{[\mu_1} \partial^{[\mu_2} h^{\nu_1] \nu_2]}$, where we antisymmetrize $\mu_{i}$ with $\nu_{i}$, but not with $\nu_j$.  In fact, all of the higher-spin field strengths in the $\rho=0$ theory can be grouped into a single object
\be
R(\xi,x)= \int d^4 \eta \delta'(\eta^2+1) e^{\xi_{\mu\nu}\eta^{\mu}\partial^{\nu}}\Psi(\eta,x),\label{highCurvatures}
\ee
where $\xi_{\mu\nu}$ is an anti-symmetric matrix variable. 
The $n$'th Taylor component of $R$ is precisely the rank $2n$ mixed-symmetry field strength for the rank-$n$ gauge field.  

Are there analogous gauge invariant, and hence possibly physical, local objects in the theory with $\rho \neq 0$?

It is straightforward to show that the \emph{only} local and covariant gauge-invariant objects in the CSP theory involving two or fewer derivatives of $\Psi(\eta,x)$ are equivalent to the equations of motion; here, we just sketch the general strategy.  Consider a completely general function $F(\xi,x) = \int [d^4\eta] \delta'(\eta^2+1) f(\eta,\xi, \partial_x) \Psi(\eta,x)$ whose gauge variation is $\delta F(\xi,x) =  \int [d^4\eta] \delta(\eta^2+1)  (-\partial_{\eta\mu} + \rho) f(\eta,\xi, \partial_x)\partial_x^{\mu}\epsilon(\eta,x))$.  The argument $\xi$ is inserted to allow completely arbitrary Lorentz structure of the invariant $F$, and we only require of $f$ that it should be decomposable as $f_0(\eta,\xi) + f^\mu_1(\eta,\xi) \partial_\mu + f^{\mu\nu}_2(\eta,\xi) \partial_\mu \partial_\nu$.  Expanding the gauge-invariance condition $\delta F(\xi,x) = 0$ in powers of $\partial_x$ gives constraints on the $f_i$, which can be solved by $f_i(\eta,\xi) = (\eta^2+1) g_i(\eta,\xi) + \dots$, where $g_i$ is an arbitrary function and the $\dots$ terms involve lower-rank $g_j$ with $j<i$.  The $O(\partial_x^3)$ component of $\delta F(\xi,x) = 0$ is simply $\partial_\eta.\partial_x f_2^{\mu\nu} = 0$. This does, of course, have solutions --- any integral of the form $\int [d^4\eta] a(\eta,\xi) E(\eta,x)$, where $E(\eta,x)$ is the left-hand side of the CSP equation of motion \eqref{eomUnfixed} and $a$ is arbitrary will be gauge-invariant by virtue of the gauge-invariance of the equation of motion.  But these objects necessarily vanish on radiation, unlike the field-strengths!  In fact, the most general solution of the gauge-invariance conditions (at quadratic order in $x$-derivatives) is equivalent to the equation of motion, integrated against the function $a(\eta,\xi) = \frac{1}{\rho^2} g_0(\eta,\xi) + \frac{1}{\rho} (\eta^2+1) \eta\cdot g_1(\eta,\xi)$. This shows that all gauge-invariant functionals of $\Psi$ satisfying our assumption of locality with only two $x$-derivatives are trivial on the support of the equations of motion.
We have not carried out this argument to higher orders in $\partial_x$, but it seems quite likely that it continues to hold if we allow $f$ to have \emph{any} finite number of $x$-derivatives.

This is not entirely surprising, since a CSP's helicity states cannot be invariantly distinguished and the high-helicity modes in \eqref{highCurvatures} require increasingly high powers of $\partial_x$ to build gauge-invariants.  On the other hand, it raises a basic question: if CSP theories are continuously connected to familiar gauge theories, then what has happened to the linear level gauge-invariants $\phi^{(0)}$, $F^{\mu\nu}$, $R^{\mu\nu\rho\sigma}$, etc.?
It is possible to build gauge-invariants that approach these in the $\rho \rightarrow 0$ limit, but the covariant ones are spatially and time-non-local functions of fields.  We can invent such an operator, but to do so requires that we introduce an un-physical auxiliary object. 
For example, if we introduce another 4-vector $\omega^\mu$, we can construct invariants
\bea
\phi(\omega,x)& =& \int d^4 \eta d^4 y \delta'(\eta^2+1) f(\omega,\eta,x-y) \Psi(\eta,y),\nonumber \\
F^{\mu\nu}(\omega,x)& =& \int d^4 \eta d^4 y \delta'(\eta^2+1) f(\omega,\eta,x-y) \eta^{[\mu}\partial^{\nu]} \Psi(\eta,y), \dots \label{nonLocalInvariants2}
\eea
with 
\be
f(\omega,\eta,x-y) = \int d^4 k e^{ik\cdot (x-y)}e^{\rho\frac{\eta\cdot\omega}{k\cdot\omega}}.
\ee
Because they are so non-local, it is completely unsurprising that $[\phi(\omega,x),\phi(\omega',x')]$ does not vanish for spacelike-separated $x-x'$, except for special  $\omega$ and $\omega'$ such as $\omega=\omega'$. Note that $\omega$ does not package the spin content of the CSP the way that vector superspace does, these functions are not analytic in $\omega$, and it's not clear what physical meaning the $\omega$-space (or spinor analogues of it) has, if any. However, the fact that such covariant operators do not simply commute for space-like separated $x-x'$ for {\it arbitrary} $\omega$ and $\omega'$
was the cause for concern flagged by several earlier author \cite{Yngvason:1970fy,Iverson:1971hq,Abbott:1976bb,Hirata:1977ss,Schroer:2013mna,Banks}. 

If we relax the requirement of manifest Lorentz covariance, we can construct gauge invariant objects that are at least local in time. 
The existence of such objects is a prerequisite for setting up the Cauchy problem in an interacting theory, starting only from physical initial data. 
If we single out a time direction, we can define 
 \be
f(\eta,\vec x-\vec y) = \int d^3 \vec k e^{-i\vec k\cdot (\vec x-\vec y)}e^{\rho\frac{\vec\eta\cdot\vec k}{\vec k ^2}}
\ee
and then construct non-covariant but time-local and gauge-invariant ``curvature'' tensors 
\bea
\phi(\vec x, t) &\equiv& \int  d^3 {\vec y} [d^4 \eta] \delta'(\eta^2+1) f(\eta,\vec x-\vec y) \Psi(\eta,{\vec y},t),\\
F^{\mu\nu}(\vec x, t) &\equiv& \int d^3 {\vec y} [d^4 \eta] \delta'(\eta^2+1) f(\eta,\vec x-\vec y) \eta^{[\mu}\partial^{\nu]} \Psi(\eta,{\vec y},t),\\
R(\xi,\vec x,t) &=& \int d^3\vec{y}[d^4 \eta] \delta'(\eta^2+1) e^{\xi_{\mu\nu}\eta^{\mu}\partial_x^{\nu}}f(\eta,\vec x- \vec y)\Psi(\eta,\vec y,t),
\eea
where the last line generalizes the series of curvatures in \eqref{highCurvatures}.  
We have explicitly separated the time and spatial dependence to underscore that the objects above are spatially non-local 
functions of $\Psi$, but local in time. 
In the $\rho\rightarrow 0$ limit, $f(\eta,\vec x-\vec y)\rightarrow \delta^3(\vec x-\vec y)$ 
so that we recover the familiar curvature forms in that limit.  Moreover, even though these are spatially non-local we nevertheless find $[R(\xi,\vec x,t), R(\xi',\vec x',t')] = 0$ for spatially separated $(\vec x,t)$ and $(\vec x',t')$, and arbitrary $\xi,\xi'$ by using the commutator \eqref{eq:commutator}.

In closing this section, we should underscore that the generalized curvatures considered here are simply motivated by their relation to familiar field strengths --- even though they obey nice causality properties, they are not especially natural objects in the vector superspace formalism, and may not be physically significant at all in the context of CSP interactions.  This is an important open question.

\section{Discussion and Conclusion}\label{Sec:conclusion}

In this paper, we have presented a Lorentz-invariant, local action for a free continuous-spin degree of freedom.  The action naturally generalizes to allow couplings of the CSP to a suitably conserved background current.   The continuity condition allows a tower of tensor currents $J^{(n)}$, of which only the lowest-rank component is conserved; this structure is compatible with no-go theorems against high-rank conserved currents in 3+1-dimensional Minkowski space, and with the covariant soft factors found in \cite{CSP1,CSP2}.
Like other local field theories for massless particles, it is a gauge theory; the canonical field from which the action and equations of motion are built is not directly observable.  This feature is a key difference between our theory and various failed attempts in the literature \cite{Yngvason:1970fy,Iverson:1971hq,Chakrabarti:1971rz,Abbott:1976bb,Hirata:1977ss}. 

Although the action is superficially similar to the one considered in our earlier work \cite{CSP3}, the new action is superior in two dramatic ways: First, the action \eqref{DdimActionCSP} propagates a single species of CSP (with arbitrary value of the Pauli-Lubanski spin invariant $\rho$), whereas the theory in \cite{CSP3} propagated a continuum of CSPs with \emph{every} $\rho$.  Second, the $\rho \rightarrow 0$ limit of this theory exhibits the expected helicity correspondence \cite{CSP2} in a much sharper way than \cite{CSP3} --- it becomes a direct sum of Schwinger-Fronsdal actions for integer-helicity particles, of which the helicity 0, 1, and 2 modes can consistently couple to the usual conserved currents.  

These two advances are made possible by the introduction of a new kind of auxiliary space, which we call ``vector superspace''.  
 Vector superspace simply unifies all integer-helicity actions in flat space, with a straightforward generalization to a CSP with non-zero spin-scale $\rho$. 
Vector superspace starts from the old idea of grouping tensors of different ranks by contracting them into an auxilliary four-vector $\eta^\mu$, then viewing their sum as a Taylor series for a function of $x$ and $\eta$.  We take this basic idea further, showing that spin can be encoded by geometry in $\eta$-space, and that integrals localized on invariant $\eta$-space hyperboloids can be unambiguously defined by analytic continuation of $\eta$.  
For ordinary ($\rho =0$) integer-helicity particles, the vector superspace action is an almost trivial rewriting of a sum of Schwinger-Fronsdal actions, though simplified and formulated in terms of {\it unconstrained} fields and gauge variations.  A simple relevant deformation of the integer-helicity action in vector superspace, which deforms the gauge invariance rather than breaking it, turns the tower of helicities into a CSP with non-zero $\rho$.  In this case, the vector superspace form is much simpler than its tensor equivalent --- the tensor action and equations of motion are not rank-diagonal, but feature rank-mixing terms of $O(\rho)$; likewise, even modes that are helicity eigenstates involve non-zero tensor components of arbitrarily high ranks.

The vector superspace action has a straightforward generalization to higher and lower dimensions, but the resulting representation content has not been thoroughly explored.  In the 2+1-dimensional case the CSP has just two degrees of freedom related by parity, which to our knowledge has not been discussed previously in the literature --- it may be possible for such degrees of freedom to arise in gapless condensed matter systems, and they display an intriguing connection to anyons at the kinematical level.  The action likely has simple generalizations to describe fermionic and/or supersymmetric CSPs, which have not yet been explored.  Another interesting open question is whether Fronsdal-type equations for massless particles in deSitter or anti-deSitter spacetimes can be recast in a vector superspace, and whether those theories admit any CSP generalization.  

\subsection*{In Search of a Covariance Principle}

A few further comments --- and a good deal more exploration --- is called for on the questions of locality, causality, and 
the general form that we might expect an interacting CSP theory to take. 

Past work has presented general arguments against covariant, single-particle gauge invariant operators with micro-causal commutation relations \cite{Yngvason:1970fy,Iverson:1971hq,Abbott:1976bb,Hirata:1977ss,Schroer:2013mna,Banks}. 
Yet here, we have a local gauge theory action where covariantly quantized fields (see \cite{CSPjenny}) have standard, causal commutation relations. 
These two findings are, of course, not incompatible, since our fields have non-trivial gauge variation and are therefore not directly observable.  Indeed, one can show that no local, single-field gauge-invariant operators with two or fewer derivatives exist, except those that are proportional to the CSP equation of motion and therefore vanish on physical states.  Though we have not carried out the general analysis explicitly, this trend likely continues to any finite order in derivatives, and is consistent with the absence of gauge-invariants with fewer than $h$ derivatives for helicity-$h$ degrees of freedom, even for $\rho=0$.  

Covariant GI operators \eqref{nonLocalInvariants2} in our theory, which Lorentz-transform like the operators considered in earlier work, can be constructed at the price of introducing non-locality in both space and time. Moreover, these are not natural objects to write down in vector superspace, and likely have no physical meaning. So it is hardly surprising that they do not have standard commutation relations.  We can, however, construct gauge-invariant operators that are local in time (but still non-local in space), at the expense of manifest Lorentz covariance.  Remarkably, the operators in this class become local in the $\rho\rightarrow 0$ limit, and commute at arbitrary spacelike separations even for non-zero $\rho$\cite{CSPjenny}!  Nonetheless, they too are somewhat artificial in vector superspace, and may well not be particularly relevant to an interacting theory.

Interpreted from the point of view of our free theory, these findings offer a clue. It seems likely that no physical role is played by operators linear in the gauge fields that are {\it invariant} under the gauge transformations.  This would have strong precedent in non-Abelian gauge theories and general relativity, which also have no local single-field gauge-invariant operators (and in GR, no local gauge-invariant operators at all).  Instead, physics in those theories is \emph{covariant} under general coordinate and gauge transformations.
For example, the equation of motion for matter in GR is not gauge-invariant even at linear level, and this non-invariance reflects the covariant transformation law of the space-time coordinates used to describe the physics.  This is to be contrasted with QED for example, where the simplest interactions exploit the existence of gauge invariant electric and magnetic fields, and no broader covariance principle is needed. 

Likewise, we should perhaps expect that the gauge-invariance of the CSP action should be interpreted as a linearized version of some coordinate transformation --- perhaps one that acts on the vector superspace (or some generalization thereof), as well as coordinate space.  Understanding consistent CSP self-interactions or couplings to matter is the next important step towards finding such a covariance principle for CSPs.  It may be fruitful to ``bootstrap'' up such interactions by expanding both in coupling and in $\rho$, starting from familiar self-interacting theories.  
We expect that the free theory formulated in this paper will be a useful stepping stone in that direction.

\section*{Acknowledgements}
The authors thank Nima Arkani-Hamed, Xavier Bekaert, and Jihad Mourad for illuminating discussions during early stages of this work.  This research was supported in part by Perimeter Institute for Theoretical Physics. Research at Perimeter Institute is supported by the Government of Canada through Industry Canada and by the Province of Ontario through the Ministry of Research and Innovation.

\appendix


\section{Definition and Examples of Regulated Vector Superspace Integrals}\label{App:tensorIntegrals}
In Section \ref{Sec:superspace}, we noted that integrals of the form $\int d^D\eta \delta(\eta^2+1) F(\eta)$ diverge for generic smooth functions $F(\eta)$ (and even for simple cases, e.g. $F(eta)=\mbox{constant}$) because the hyperboloid on which the integral is evaluated has infinite volume.  
Nonetheless, these integrals can be computed either by analytic continuation in $\eta$-space (which we take as a definition) or equivalently, up to normalization, directly from their symmetry properties in Minkowski space.  This Appendix carefully defines these measures and presents useful results and examples.  \S\ref{Sapp:anacont} defines the measure $[d^D\eta]$ using analytic continuation and derives generating functions for integrals over both $\delta(\eta^2+1)$ and $\delta'(\eta^2+1)$.  In \S\ref{Sapp:minkowskiTensor}, we explain how the same integrals could alternately be derived (up to an overall normalization) using symmetries and integration by parts directly in Minkowski space, and present a simple example.  As a further illustrative example of $\eta$-space computations, \S\ref{Sapp:orthonormality} demonstrates the orthonormality of the basis wavefunctions for CSPs introduced in \S\ref{Ssec:planewave}.

\subsection{Regulated Measure over Vector Superspace from Analytic Continuation}\label{Sapp:anacont}
For any smooth function $F(\eta)$ we define $\eta$-integration over the hyperboloid $\eta^2+1  = 0$ via analytic continuation.  Specifically, 
\be
\int [d^D \eta] \delta(\eta^2+ 1) F(\eta) \equiv \frac{2}{S_{D-1}} \int d^D \bar\eta \delta(\bar\eta^2 - 1) F(\bar\eta),\label{defWick}
\ee
where $\bar\eta^\mu = (i\eta^0, \eta^1, \dots \eta^{D-1})$ are Wick-rotated coordinates, $F(\bar\eta)$ is defined by analytic continuation from real $\eta$, and $S_{D-1}$ is the surface-area of a unit $(D-1)$-sphere, so that with the normalization above $\int [d^D\eta] \delta(\eta^2+1) = 1$.  Similarly, integrals over a $\delta'$ are given by 
\be
\int [d^D \eta] \delta'(\eta^2+ 1) F(\eta) \equiv \frac{2}{S_{D-1}} \int d^D \bar\eta \delta'(-\bar\eta^2 + 1) F(\bar\eta) =  - \frac{1}{S_{D-1}} \int d^D \bar\eta \delta'(\bar\eta^2 - 1) F(\bar\eta),\label{defWickPrime}
\ee
where we have used $\delta'(-x) = -\delta(x)$.   The above expressions are clearly finite for any smooth $F$, since they are integrals of smooth functions over a compact surface.  

In fact, since any smooth $F(\eta) = \left[F(i \partial_w) e^{i\eta.w} \right]_{w=0}$, we can rewrite the Wick-rotated integrals \eqref{defWick} and \eqref{defWickPrime} as 
\bea
\int [d^D \eta] \delta(\eta^2+ 1) F(\eta) & = & \left[ F(i \partial_w) G(w)\right]_{w=0} = \left[ G(i \partial_\eta) F(\eta)\right]_{\eta=0} \equiv F*G  \label{FstarG} \\ 
\int [d^D \eta] \delta'(\eta^2+ 1) F(\eta) & =& \left[ F(i \partial_w) G'(w)\right]_{w=0} = \left[ G'(i \partial_\eta) F(\eta)\right]_{\eta=0} \equiv F*G' \label{FstarG'}
\eea
where $F*G$, the Moyal star product, is a common short-hand for the preceding expressions.  The generating functions $G$ and $G'$ are defined via analytic continuation as 
\bea
G(w) &\equiv & \frac{2}{S_{D-1}} \int d^D \bar\eta \delta(\bar\eta^2 - 1) e^{-i\bar\eta.\bar w} \bigg|_{\bar w = {iw^0,w^i}} =  \Gamma(\tfrac{D}{2}) (r/2)^{-\tfrac{D-2}{2}} J_{\tfrac{D-2}{2}}(r)\bigg|_{r=\sqrt{-w^2}}\label{Gfunc}\\
G'(w) &\equiv &  \frac{2}{S_{D-1}} \int d^D \bar\eta \delta'(\bar\eta^2 - 1) e^{-i\bar\eta.\bar w}|_{\bar w = {iw^0,w^i}} = \Gamma(\tfrac{D}{2}) (r/2)^{-\tfrac{D-4}{2}} J_{\tfrac{D-4}{2}}(r)\bigg|_{r=\sqrt{-w^2}}. \label{Gpfunc}
\eea
That these are built from Bessel functions is not surprising, since the condition 
$\int d^4\bar\eta \delta(\eta^2 + 1) (\eta^2 + 1) e^{-i\eta.w}$ is equivalent to $(\partial_w^2 - 1)G(w)= 0$.
Using the invariance of $G$ under Lorentz transformations of $w$, we can write $G(w)=G(r)$ with $r=\sqrt{-w^2}$, this implies $d^2G/dr^2 + \frac{D-1}{r} dG/dr - G = 0$.  This is precisely the $D$-dimensional analogue of the Bessel equation!  

For $D=4$ and $D=3$, the general formulas above reduce to 
\begin{align}
&G_{4d}(r)=2 J_1(r)/r \quad &G'_{4d}(r) =& J_0(r) \\
&G_{3d}(r) = j_0(r) = \sin(r)/r \quad& G'_{3d}(r) =& \cos(r)/2.
\end{align}
Thus, for example, we can easily verify that in 4 space-time dimensions 
\bea
\int [d^4\eta] \delta'(\eta^2+1) \eta^\mu \eta^\nu A_{\mu\nu} & = & \left[ A_{\mu\nu} (i \partial_w^\mu) (i \partial_w^\nu) J_0(\sqrt{-w^2})\right]_{w=0}\\
 & = & A_{\mu\nu} \left[-g^{\mu\nu} J_1(r)/r + w^\mu w^\nu J_2(r)/r^2 \right]_{w=0} = - A^\mu_\mu/2. \label{eq:integrationExample1}
 \eea

With \eqref{Gfunc} and \eqref{Gpfunc} in hand, we can re-express the action \eqref{DdimAction} as 
\be
S  =  \left[ G(i\partial_\eta) \frac{1}{2}(\partial_\alpha \Psi)^2 - \frac{1}{4} (\eta^2+1) (\Delta \Psi)^2 \right]_{\eta=0}. \label{StarActionApp}
\ee

The qualitative resemblance of our $\eta$-space to superspace (where integration in the fermonic coordinate $\theta$ can also be re-expressed as differentiation, there owing to the Grassmann nature of $\theta$) is apparent.  But here, because $\eta$ is bosonic, we are led to include an \emph{infinite} tower of same-statistics fields rather than a finite tower of fields with alternating statistics.

\subsection{Vector Superspace Integrals from Symmetry Arguments}\label{Sapp:minkowskiTensor}
The regulated $\eta$-space measure defined by \eqref{defWick} respects the $\delta$-function and integration-by-parts identities 
\be
\int [d^D \eta] \delta(\eta^2 + 1) (\eta^2+1) F(\eta) = 0 \qquad \int [d^D \eta] \partial_\eta^\mu \left( \delta(\eta^2 + 1) F(\eta) \right) = 0, 
\ee
which in turn guarantee 
\be
\int [d^D \eta] \delta'(\eta^2 + 1) F(\eta) = \int [d^4 \eta] \delta(\eta^2 + 1) \tfrac{1}{2} (D-2 + \eta\cdot\partial_\eta) F(\eta). \label{deltaPrime}
\ee
In fact, we can essentially \emph{derive} the correct measure, up to an overall normalization, from these identities.   To take a concrete example, if we define a constant $V$ by  
\be
V \equiv \int [d^4 \eta] \delta(\eta^2 + 1), 
\ee
then by symmetry
\bea
\int [d^4 \eta] \delta'(\eta^2 + 1) \eta^\mu \eta^\nu &= &\int [d^4 \eta] \delta(\eta^2 + 1) (1+\tfrac{1}{2} \eta \cdot \partial_\eta ) \eta^\mu \eta^\nu \nonumber \\
& = & 2 \int [d^4 \eta] \delta(\eta^2 + 1) \eta^\mu \eta^\nu =  \frac{1}{2}\int [d^4 \eta] \delta(\eta^2 + 1) \eta^2 g^{\mu\nu} \nonumber \\
&= & \frac{1}{2} g^{\mu\nu} \int [d^4 \eta] \delta(\eta^2 + 1) (-1) = -\frac{1}{2} g^{\mu\nu} V,\label{eq:integrationExample2}
\eea
which is readily seen to agree with the result of \eqref{eq:integrationExample1} when we adopt the normalization $V=1$.  

The style of argument in \eqref{eq:integrationExample2} is readily generalized to relate the rank-$n$ tensor $\int [d^4\eta] \delta(\eta^2+1) \eta^n$ to $\int
 [d^4\eta] \delta(\eta^2+1) \eta^{n-2}$ and similarly for the integrals over $\delta'$.  The results of this argument precisely reproduce what is expected from a series expansion of \eqref{FstarG}-\eqref{Gpfunc}.
\subsection{Orthonormality of basis wave-functions}\label{Sapp:orthonormality}

As an example of vector superspace computations, we explicitly compute the orthonormality relation
\be
\braket{\psi_h(k)}{\psi_{h'}(k)} \equiv \int [d^4\eta] \delta'(\eta^2+1) \psi_h(\eta,k)^* \psi_{h'}(\eta,k) = \delta_{h,h'} \label{eq:TheIntegral}.
\ee
for the basis wave-functions
\be
\psi_{+h}(\eta,k) = (\eta\cdot \epsilon_+)^{h}e^{i\rho\eta\cdot q} \quad \psi_{-h}(\eta,k) = (\eta\cdot \epsilon_-)^{|h|}e^{i\rho\eta\cdot q}.
\ee
Upon analytic continuation, we can choose a coordinate system with 
$$\eta_E^{\mu}\equiv(\eta_{\perp}\cos\theta,\eta_{\perp}\sin\theta,\tilde{\eta}\cos\phi,\tilde{\eta}\sin\phi),$$
where $\eta_{\perp}^2 = \eta\cdot \epsilon_+\eta\cdot \epsilon_-$ labels the magnitude of $\eta_E$ in the plane spanned by $\epsilon_{\pm}$.
In this coordinate system, we can write $\eta\cdot \epsilon_+=\eta_{\perp}e^{i\theta}$. 
Using $[d^4\eta]=\frac{1}{\pi^2}d^4\eta_E$, we can write the integral measure as $\frac{1}{\pi^2}\eta_{\perp}d\eta_{\perp}\tilde{\eta}d\tilde{\eta}d\theta d\phi$
and easily evaluate \eqref{eq:TheIntegral} as
\bea
 \int [d^4\eta] \delta'(\eta^2+1) \psi_h(\eta,k)^* \psi_{h'}(\eta,k) &=& 
-\frac{1}{\pi^2}\int \eta_{\perp}d\eta_{\perp}\tilde{\eta}d\tilde{\eta}d\theta d\phi \delta^{\prime}(\eta_{\perp}^2+\tilde{\eta}^2-1) \eta_{\perp}^{|h|+|h'|}e^{(h'-h)\theta} \nonumber \\
&=& -\delta_{h,h'}\frac{1}{\pi^2}\int \eta_{\perp}d\eta_{\perp}\tilde{\eta}d\tilde{\eta}d\theta d\phi \delta^{\prime}(\eta_{\perp}^2+\tilde{\eta}^2-1) \eta_{\perp}^{2h} \nonumber \\
&=& -\delta_{h,h'}\int_0^{\infty}du \int_0^{\infty} \delta^{\prime}(u+v-1) u^h \nonumber \\
&=& \delta_{h,h'}\int_0^{\infty} du \delta(u-1) u^h = \delta_{h,h'}.
\eea
Similar calculations in vector superspace are carried out in the same manner. 

 \section{Diagonalizing the Vector Superspace Action Without Constraints}\label{App:unconstrainedRankDiagonal}
The results of the preceding section make clear that, if we expand $\Psi$ about $\eta=0$ as $\Psi(\eta,x) = \psi^{(0)}(x) + \eta^\mu \psi^{(1)}_\mu(x) + \eta^\mu \eta^\nu \psi^{(2)}_{\mu\nu}(x) + \dots$, with the tensor fields $\psi^{(n)}$ unconstrained, the action \eqref{StarActionApp} will mix $\psi^{(n)}$ with $\Tr^t \psi^{(n+2t)}$ for all integer $t$, even for $\rho=0$.   

This rank-mixing is related to the fact that we defined our polynomials via Taylor expansion about $\eta=0$, but in fact the action is evaluated at non-zero $\eta$.  It can be simply removed by a (non-unique) change of variables.  We have shown already in \S\ref{Ssec:tensorDecomposition} that an appropriate decomposition of $\Psi$ in terms of tensor fields \eqref{FronsdalDecomposition} yields a rank-diagonal action, provided we first partially fix gauge so that the tensor fields $\phi^{(n)}$ are double-traceless.  It is, however, possible to construct tensor fields from $\Psi$ such that the action is both simple and rank-diagonal \emph{before any gauge-fixing}.  

Concretely, we can introduce unconstrained tensor fields
\bea
A^{(n)}(x)  & = & \sum_{t=0}^{\infty} \frac{(-1)^t (2 n+D-4)!! (2t+n)!}{2^{n/2+t} (2n+2t+D-4)!! t! n!}Tr^t \psi^{(n+2 t)}(x) \\
& = & \sum_{t=0}^{\infty} \frac{(-1)^t (2 n+D-4)!! (\partial_\eta^2)t \partial_\eta^{\mu_1} \dots \partial_\eta^{\mu_n}}{2^t (2n+2t+D-4)!! t! n!}\Psi(\eta,x) \big|_{\eta = 0}\label{eq:TensorFields}
\eea
in $D$ space-time dimensions.  For $D=4$, these decompositions simplify to
\bea
A^{(n)}_{4d}(x) & = & \sum_{t=0}^{\infty} \frac{(-\partial_\eta^2/4)^t \partial_\eta^{\mu_1} \dots \partial_\eta^{\mu_n}}{2^{n/2} t! (t+n)!}\Psi(\eta,x) |_{\eta = 0},
\eea
which is precisely the Taylor expansion of $\left[J_n(\sqrt{\partial_\eta^2}) \Psi(\eta,x)\right]_{\eta=0}$ up to the normalization and additional Lorentz index-structure.   In terms of these new fields, the $\rho=0$ action \eqref{StarAction} becomes rank-diagonal 
\be
S  = \sum_n S_n
\ee
where (for $D=4$) 
\be
S_n = (-1)^n \int d^4 x \frac{1}{2} (\partial_\alpha \tsub{A^{(n)}})^2 - \frac{1}{8} (\partial_\alpha \tsub{{A^{(n)}}'})^2 - \frac{n}{2} \tsub{\partial\cdot A^{(n)}}^2, \label{eq:action}
\ee
where $A' = \Tr A$ and $\tsub{A}$ denotes the ``trace-subtraction'' of $A$ --- the \emph{unique} fully symmetric, traceless rank-$n$ tensor built from only $A$ itself. For example, in 3+1 dimensions we have $\tsub{A}^{\mu\nu} \equiv A^{\mu\nu} - \frac{1}{4} g^{\mu\nu} A'$, $\tsub{A}^{\mu\nu\rho} \equiv A^{\mu\nu\rho} - \frac{1}{6} (g^{\mu\nu} A'^\rho+g^{\rho\mu} A'^\nu+g^{\nu\rho} A'^\mu )$ and so on, with the general form 
\be
\tsub{X} \equiv X  + \sum_{t=1}^{n/2} \frac{(-1)^t (n-t)!}{2^t n!} \eta^{(t)} \circ  \Tr^t X. \label{tsubdef}
\ee

Appropriate decompositions of  $\varepsilon(\eta,x)$ and $\chi(\eta,x)$ into components $\varepsilon^{(n)}$ and $\chi^{(n)}$ relates the $\eta$-space gauge variations \ref{eq:fullGaugeRedundancy} to component gauge variations
\be
\delta A^{(n)} = \partial \circ \varepsilon^{(n-1)} - g \circ (\partial\cdot\varepsilon^{(n-1)}) + g \circ g \circ \chi^{(n-4)} \label{eq:gauge},
\ee
which are easily seen to leave the action \eqref{eq:action} invariant.  
The gauge-invariant equation of motion is
\be
-\Box\tsub{A} + \frac{1}{2 s(s-1)} \eta \circ \Box \tsub{A'} +\partial \circ \tsub{\partial\cdot A}= 0. \label{eq:eom}
\ee
This is readily seen by using the tracelessness of $\tsub{\dots}$, which implies  
$\tsub{\delta A} .\tsub{A} = \delta A . \tsub{A}$ (since $\tsub{\delta A}- \delta A$ is proportional to the metric, 
and hence annihilates $\tsub{A}$ upon contraction), and similarly for the other two terms in the variation of the action.  

As was noted earlier, the gauge transformation \eqref{eq:gauge} is redundant because any transformation 
generated by $\chi$ can also be generated by $\epsilon=g\circ \xi$ with $\partial\cdot\xi=-\chi$.  In other words, we can parametrize an arbitrary gauge variation \emph{either} by $\epsilon$ alone (with $\chi=0$) or by a \emph{traceless} $\epsilon$ and an arbitrary $\chi$.  

 We can again check that we have only two massless propagating degrees of freedom permitted by this system of equations 
subject to the gauge redundancies generated by $\tsub{\epsilon}$. 
A double traceless rank-n field has $\frac{1}{6}[(n+3)(n+2)(n+1)-(n-1)(n-2)(n-3)]$ independent components. 
The constraint $\tsub{\partial\cdot A} = 0$ and the remaining gauge parameters $\tsub{\epsilon}$ both have 
$\frac{n}{6}[(n+2)(n+1)-(n-1)(n-2)]$ independent components. The number of field components minus the number of 
constraints and gauge parameters is therefore 2, as we'd expect for a massless helicity-n degree of freedom. 
 
This action can be rewritten in the Schwinger-Fronsdal form by first using the $\chi$ gauge-invariance to fix a double-traceless gauge for $A^{(n)}$, then rewriting the action in terms of a trace-reversed field  $\phi^{(n)} \equiv \tsub{A} - \frac{1}{2(s-1)}  g\circ \tsub{A'}$ (this is analogous to the gauge-fixing used in \S\ref{Ssec:tensorDecomposition}), which is readily seen to transform in the usual way under the remaining gauge variations, parametrized by traceless $\epsilon$.
Incidentally, using the relation 
\be
 \tsub{\partial.T} =  \partial.\tsub{T} +\frac{1}{2n} \left[ \partial\circ \tsub{T'} - \frac{1}{n-1} g \circ \partial.\tsub{T'}\right],\label{dTidentity}
\ee
the equation of motion \eqref{eq:eom} (the variation of the action with respect to $A^{(n)}$, not $\phi^{(n)}$) can be written as 
\be
(\Box - \partial \circ \partial \cdot + \partial^2 \circ g \cdot) \phi^{(n)} = 0,
\ee
which is precisely the Fronsdal equation for $\phi^{(n)}$.  

In the $A^{(n)}$ language, Fronsdal's double-trace condition is most naturally viewed as a partial gauge-fixing.  The restriction to traceless $\epsilon$ is replaced by the statement that trace-terms in $\epsilon$ either have trivial action on the fields $A^{(n)}$ or violate the double-trace condition.  It is, of course, well known that trace conditions can be replaced by projections in the action, which is essentially what we have done here, but in a convenient basis for making contact with the vector superspace action.

For nonzero $\rho$, the action \eqref{eq:action} changes only by the substitution 
$$\tsub{\partial.A}\rightarrow \left<\partial.A^{(n)} + \frac{\rho}{\sqrt{2} n}\big( A^{(n-1)} + \tfrac{1}{2} {A^{(n+1)}}'\big)\right>,$$
with corresponding modification to both the gauge transformation \eqref{eq:gauge} and the equation of motion \eqref{eq:eom}.

\bibliography{SingleCSPGaugeTheory}
\end{document}